# Defect Passivation of 2D Semiconductors by Fixating Chemisorbed Oxygen Molecules via *h*-BN Encapsulations

*Jin-Woo Jung, Hyeon-Seo Choi, Young-Jun Lee, Youngjae Kim, Takashi Taniguchi, Kenji Watanabe, Min-Yeong Choi, Jae Hyuck Jang, Hee-Suk Chung, Dohun Kim, Youngwook Kim, and Chang-Hee Cho\**

Jin-Woo Jung, Hyeon-Seo Choi, Young-Jun Lee, Dohun Kim, Youngwook Kim, Chang-Hee Cho
Department of Physics and Chemistry, Daegu Gyeongbuk Institute of Science and Technology (DGIST), Daegu 42988, South Korea
E-mail: chcho@dgist.ac.kr

Youngjae Kim
School of Physics, Korea Institute for Advanced Study (KIAS), Seoul 02455, South Korea

Takashi Taniguchi
International Center for Materials Nanoarchitectonics, National Institute for Materials Science, Tsukuba 305-0044, Japan

Kenji Watanabe
Research Center for Functional Materials, National Institute for Materials Science, Tsukuba 305-0044, Japan

Min-Yeong Choi, Jae Hyuck Jang, Hee-Suk Chung
Electron Microscopy and Spectroscopy Team, Korea Basic Science Institute, Daejeon 34133, South Korea

Jae Hyuck Jang
Graduate School of Analytic Science and Technology, Chungnam National University, Daejeon 34134, South Korea

Hexagonal boron nitride ($h$-BN) is a key ingredient for various two-dimensional (2D) van der Waals heterostructure devices, but the exact role of $h$-BN encapsulation in relation to the internal defects of 2D semiconductors remains unclear. Here, we report that $h$-BN encapsulation greatly removes the defect-related gap states by stabilizing the chemisorbed oxygen molecules onto the defects of monolayer $WS_2$ crystals. Electron energy loss spectroscopy (EELS) combined with theoretical analysis clearly confirms that the oxygen molecules are chemisorbed onto the defects of $WS_2$ crystals and are fixated by $h$-BN encapsulation, with excluding a possibility of oxygen molecules trapped in bubbles or wrinkles formed at the interface between $WS_2$ and $h$-BN. Optical spectroscopic studies show that $h$-BN encapsulation prevents the desorption of oxygen molecules over various excitation and ambient conditions, resulting in a greatly lowered and stabilized free electron density in monolayer $WS_2$ crystals. This suppresses the exciton annihilation processes by two orders of magnitude compared to that of bare $WS_2$. Furthermore, the valley polarization becomes robust against the various excitation and ambient conditions in the $h$-BN encapsulated $WS_2$ crystals.

## 1. Introduction

Monolayer transition metal dichalcogenides (TMDs) have emerged as a platform to examine various exciton species such as trions, biexcitons, interlayer excitons, and moiré excitons due to the strong inter-particle interactions.[1–4] Furthermore, the valley-dependent optical selection rules given by the broken inversion symmetry enable the selective generation of excitons in the particular valley (+K or −K) using circularly polarized light,[5] providing the opportunity for applications toward valleytronic devices. However, the external disorders in close proximity of monolayer TMDs such as substrate-induced surface roughness and absorbates can significantly alter the excitonic properties of two-dimensional (2D) TMD materials, hindering the observation of the unique properties in monolayer TMDs.[6,7]

In an attempt to reduce the disorder from substrates, it has been proposed to encapsulate TMD materials using hexagonal boron nitride ($h$-BN) layers.[8,9] Recently, it has been shown that $h$-BN encapsulation enables to observe the intrinsic optical properties of monolayer TMDs, including the excitonic linewidth with homogeneous broadening limit[9] and the suppression of exciton annihilation processes.[10] As the origin, the reduced substrate disorders have often been suggested in the previous works,[9,10] but the excitonic properties of TMDs on $h$-BN substrates show large discrepancies from those of TMDs encapsulated by $h$-BN.[11,12] On the other hand, previous investigations have shown that the oxygen molecules on the surface of TMDs also alter the electronic and optical properties of 2D TMD materials.[13–18] The adsorption of the oxygen molecules occurs on the defects with relatively lower kinetic barrier rather than the perfect sites of TMD materials,[18] and the oxygen molecules unlike other molecules can only be chemisorbed at chalcogen vacancies due to isovalent valence electrons (two unpaired electrons) with the chalcogen atom.[14,15] This chemisorption of oxygen molecules at the defect sites removes the defect-related gap states without significantly altering the electronic band structures of TMD materials.[13,14] Thus, the oxygen molecules, that are supplied during the exposure of TMDs into the atmosphere, can significantly change the properties of defect-related states through the chemical adsorption process. In this regard, $h$-BN encapsulation can play a crucial role in the defect states of TMDs, in which the $h$-BN layers fixate the adsorbed oxygen molecules on the TMD defects and facilitate the interaction between the oxygen molecules and the defect states. However, the role of $h$-BN encapsulation in relation to the defects of TMDs remains unexplored.

In this work, we found that $h$-BN encapsulation stabilizes the chemisorbed oxygen molecules on the defect sites of monolayer $WS_2$ crystals, which greatly passivates the defect-related gap states along with the decrease in the free electron density. Electron energy loss spectroscopy (EELS) combined with theoretical analysis clearly reveals that the oxygen molecules are chemisorbed onto the defects of $WS_2$ crystals and are fixated by $h$-BN encapsulation, that excludes a possibility of oxygen molecules trapped in bubbles or wrinkles formed at the interface between $WS_2$ and $h$-BN. Optical spectroscopic studies show that $h$-BN encapsulation prevents the desorption of oxygen molecules over various excitation and ambient conditions, resulting in a greatly lowered and stabilized free electron density in monolayer $WS_2$ crystals. This suppresses the exciton annihilation processes by two orders of magnitude compared to that of bare $WS_2$. Furthermore, due to the stabilized free electron density in the $h$-BN encapsulated $WS_2$ crystals, the valley polarization becomes robust against the elevated excitation condition.

## 2. Results and Discussion

Figure 1a shows a schematic illustration showing that the fixated oxygen molecules by the $h$-BN layers effectively passivate defects of the $WS_2$. We considered the chemisorption type, where the oxygen molecule chemically bonds to three surrounding tungsten atoms, which is the most common configuration for the oxygen chemisorption (inset image of Figure 1a) in the S-based TMDs.[14–17] The chemisorbed oxygen molecules at the chalcogen vacancies can also be dissociated into two oxygen atoms, leading to a dissociative chemisorption, which occupies the sulfur vacancies with the dissociated oxygen atoms.[13–15] However, in the case of the $WS_2$ used in our study, the kinetic barrier for the $O_2$-chemisorption (0.56 eV) is lower than that for the $O_2$-dissociative chemisorption process (0.76 eV). It is estimated that the probability of the $O_2$-chemisorption is 1000 times higher than that of the dissociative chemisorpiton (see Figure S1 in Supporting Information). Thus, the oxygen chemisorption on the monolayer $WS_2$ crystals would have the final configuration of the $O_2$-chemisorption rather than the $O_2$-dissociative chemisorption.[14] The major molecules such as $N_2$, $O_2$, and $H_2O$ in air can be weakly physisorbed at both the pristine surface and defect sites of $WS_2$. However, this physisorption has virtually no influence on the electronic and optical properties of the $WS_2$ monolayer due to easy desorption of physisorbed molecules.[14] In addition, our first-principle calculations demonstrate that the oxygen molecules can only be chemisorbed onto the defects (sulfur vacancy) and attain a fully stable chemisorption state, indicating that the oxygen molecules can be majorly adsorbed onto the $WS_2$ in the air. The detailed theoretical calculation results on molecular interactions with the sulfur vacancy and pristine surface of $WS_2$ are provided in Supporting Information S2 and S3. To investigate the role of oxygen fixation in the excitonic properties of monolayer $WS_2$ with excluding the effects of disorders induced by the substrates, we studied *h*-BN encapsulated $WS_2$ crystals suspended on line trenches with a linewidth of 1.8 μm in comparison with bare $WS_2$, as shown in Figures 1b,d. Scanning electron microscope images confirm the suspended structures for both the bare (Figure 1b) and *h*-BN encapsulated (Figure 1d) $WS_2$ crystals on the line trenches (see Figure S4 in Supporting Information). The monolayered $WS_2$ crystals used in this study were grown on sapphire substrates using a chemical vapor deposition (CVD) method.[19] Mechanically exfoliated *h*-BN flakes with a thickness of ~40 nm were used as the encapsulating layers in the *h*-BN/$WS_2$/*h*-BN structures. The detailed sample preparation processes are described in the methods. Figures 1c,e display the spatial photoluminescence profiles of the bare (Figure 1c) and *h*-BN encapsulated (Figure 1e) $WS_2$ suspended on the line trenches, respectively. The steady-state photoluminescence measurements were carried out at a low level of excitation (~0.065 kW/cm$^2$) to rule out the heating effect. It is worth noting that the photoluminescence intensity becomes stronger in the suspended regions than in the supported regions for both the bare and *h*-BN encapsulated $WS_2$ crystals due to the enhanced local field effect by optical interference in the trench region.[7] To confirm the exciton species, the photoluminescence spectra were measured at a cryogenic temperature of 77 K under a vacuum level of ~1 × 10$^{-5}$ Torr, as shown in Figure 1f. For the *h*-BN encapsulated $WS_2$, the neutral exciton ($X^0$) and the trion ($X^-$) are identified at energies of 2.042 and 2.001 eV, respectively, while the bare $WS_2$ shows three species of the neutral exciton ($X^0$), trion ($X^-$), and defect-related trapped exciton (L) at 2.087, 2.042, and 2.026 eV, respectively. The energy of neutral exciton was assigned by measuring the differential reflectance spectra (see Figure S5 in Supporting Information), and those of trion and defect-related trapped exciton were identified by the energy differences from the neutral exciton.[20,21] Compared to the bare $WS_2$ showing the emission prevailed by the trion and the defect-related trapped exciton, the *h*-BN encapsulated $WS_2$ exhibits the predominant emission from the neutral exciton with homogeneous linewidth of the 6 meV,[9] indicating that the defect-induced free electrons and inhomogeneous broadening are substantially reduced by the encapsulating $WS_2$ with *h*-BN layers. These results indicate that the oxygen fixation by *h*-BN encapsulation can play a crucial role in the defect removal with reducing the free electron density. To directly confirm the fixation effect of the adsorbed oxygen molecules on the defects, the change in excitonic spectra was monitored under the different ambient conditions of air and vacuum, as shown in Figure 1g (see Figure S6 in Supporting Information). Striking features are observed for the bare $WS_2$ crystals, showing that the spectral weight of neutral excitons is predominant over that of trions under ambient air condition, whereas that of trions becomes larger than that of neutral excitons under vacuum. These results indicate that the oxygen adsorbates on the defect sites are released by changing the ambient condition from air to vacuum, raising the density of free electrons in the bare $WS_2$ crystals under vacuum.[16,17] As shown in Figures S7, the *h*-BN encapsulated $WS_2$ samples fabricated under an inert ($N_2$) environment exhibit much stronger trion intensity (higher free electron concentration) compared to that of the *h*-BN encapsulated $WS_2$ fabricated in the air. The spectral feature is very similar to that of the bare $WS_2$ measured in the vacuum environment (see top panels in Figures 1f and 1g). Furthermore, the exfoliated monolayer $WS_2$ and $WSe_2$ with a lower density of chalcogen vacancies give rise to a less change in the free electron density against the variation of the ambient conditions, implying that the oxygen molecules are mostly adsorbed on the chalcogen vacancies (see Figure S8 in Supporting Information). In contrast, the *h*-BN encapsulated $WS_2$ crystals exhibit almost the same spectra prevailed by the neutral excitons regardless of the ambient conditions, highlighting that the *h*-BN encapsulation effectively removes the internal defects by stabilizing the oxygen molecules adsorbed onto $WS_2$ crystals.

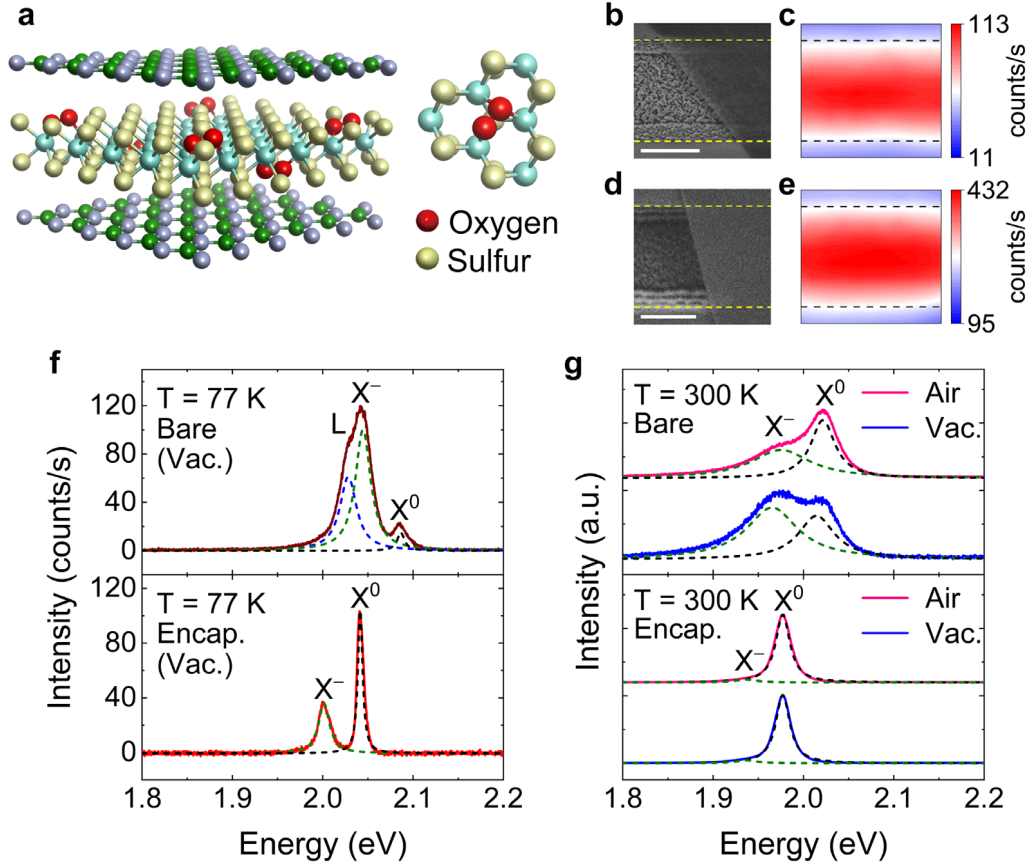

**Figure 1.** a) Schematic illustration showing the chemisorbed oxygen molecules anchored by the *h*-BN encapsulation. Right inset image represents the detailed atomic configuration of the chemisorbed oxygen molecule at the sulfur vacancy. b) Scanning electron microscope image of the bare WS$_2$ crystals on the line trenches. c) Spatial photoluminescence profile measured from the bare WS$_2$ on the line trench. d) Scanning electron microscope image of the *h*-BN encapsulated WS$_2$ crystals on the line trench. e) Spatial photoluminescence profile measured from the *h*-BN encapsulated WS$_2$ on the line trenches. The yellow and black dashed lines marked in (b,d) and (c,e) indicate the boundary of the suspended and supported regions. The scale bars of (b,d) are 1 μm. f) Photoluminescence spectra for the bare (top panel) and *h*-BN encapsulated (bottom panel) WS$_2$ measured at a cryogenic temperature of 77 K under a vacuum level of ~1 × 10$^{-5}$ Torr. g) Photoluminescence spectra for the bare (top panel) and *h*-BN encapsulated (bottom panel) WS$_2$ measured under different ambient conditions of air and vacuum (T = 300 K). Each photoluminescence spectrum was fitted by Lorentzian functions. The black, olive, and blue dashed lines represent the neutral exciton (X$^0$), trion (X$^-$), and defect-related trapped exciton (L) states, respectively.

The EELS analysis for oxygen $K$-edge confirms that the $h$-BN encapsulation anchors the chemisorbed oxygen molecules onto the defects of monolayer $WS_2$ crystals. Figure 2a displays the EELS spectra of the oxygen $K$-edge for the $h$-BN encapsulated $WS_2$, bare $WS_2$, and $h$-BN flake crystals. The $h$-BN encapsulated $WS_2$ crystals show the oxygen $K$-edge peaks centered at 538 and 556 eV, whereas any oxygen-related features are not observed for the bare $WS_2$ and the $h$-BN crystals. This suggests that the adsorbed oxygen molecules on $WS_2$ are fixated by the $h$-BN encapsulation (see Figure S9 in Supporting Information). Figures 2b-e show the EELS maps for the boron (b), nitrogen (c), oxygen (d) $K$-edge, and the sulfur (e) $L$-edge meausred from the $h$-BN encapsulated $WS_2$. Note that the elemental signal displayed for a pixel in the EELS map is the sum of the elemental signals detected through 2D scanning using an electron beam with a spatial resolution of 1 Å over an area of 20 nm × 20 nm. The spatial EELS maps confirm that the oxygen $K$-edge signal is markedly weak compared to those of other elements, and is randomly distributed in the 2D plane of $h$-BN encapsulated $WS_2$ samples. This strongly indicates that the oxygen molecules are chemisorbed on the randomly distributed local defects in $WS_2$ crystals (see Figure S10 in Supporting Information).

To discover the adsorption types of the adsorbed oxygen molecules by $h$-BN encapsulation, we demonstrate the theoretical EELS spectra of oxygen $K$-edge for the physisorbed oxygen molecule on the pristine $WS_2$ (Figure 2f) and chemisorbed oxygen molecule at the sulfur vacancy site of the $WS_2$ (Figure 2g). To understand the underlying physics for the EELS results, we performed theoretical EELS calculations based on the first-principles calculations implemented in the full-potential linearized plane wave (FLAPW) + local orbitals with the ELK code. Here, we used the pseudo core-hole method, where the self-consistent calculation is made in terms of one of the oxygen nuclei constrained to be positively charged (+1e) and an additional electron (−1e) is simultaneously constrained to occupy the conduction orbitals. After the self-consistent calculation, the Kohn-Sham orbitals become well defined, and then we compute the photon-absorption matrix elements between the core orbital ($s$-orbital) of the oxygen nuclei and the unoccupied conduction bands, which corresponds to the dielectric function for EELS spectra (see Supporting Information S11 for more detail).

The calculated EELS spectrum for the physisorbed oxygen molecule on the pristine $WS_2$ indicates the two main peaks at 530 eV and 539 eV, which originate from the transition of the core electrons to two kinds of trivial hybridized states in the oxygen molecule, featured as anti-bonding orbital $\pi^*$ (top) and $\sigma^*$ (bottom) distributions, respectively (see the inset of Figure 2f), whereas the calculated oxygen $K$-edge peaks for the chemisorbed oxygen molecule appear at the energy loss positions of 538 and 555 eV (labelled as A and B on the spectrum), respectively.

Since the core-hole excitation makes the oxygen molecule to have an asymmetric potential, the $\pi^*$ orbital distribution in the inset reflects asymmetric densities. As shown in Figure 2g, the calculated EELS spectrum (red dashed) for a chemisorbed oxygen molecule is in good agreement with that of the $h$-BN encapsulated $WS_2$ (blue line), indicating that the fixated oxygen molecules by the $h$-BN encapsulation are chemisorbed at the defect sites of the $WS_2$. The anti-bonding orbital $\pi^*$ peak (530 eV) is absent for the chemisorbed oxygen molecule in the calculated EELS spectrum. The disappearance of $\pi$-bonding is commonly interpreted as a transformation of bonding sequences.[22] In our study, it is found that the hybridizations between the oxygen molecule and the surrounding tungsten atoms directly suppress the $\pi^*$ peak. As shown in the top inset of Figure 2g, the real-space orbital distribution for the peak A resembles an orbital shape of the σ* peak for the physisorbed oxygen molecule, indicating that the hybridizations between the oxygen molecule and the tungsten atoms induce σ* antibonding energy state similar to that of the physisorbed oxygen molecule. In contrast, for the peak B, the real-space orbitals show highly delocalized distribution for both the oxygen molecule and the $WS_2$ crystal (the bottom inset of Figure 2g), implying that the peak B is due to the transition of the core electrons to continuum bands contributed from both the oxygen molecules and the $WS_2$ crystals.

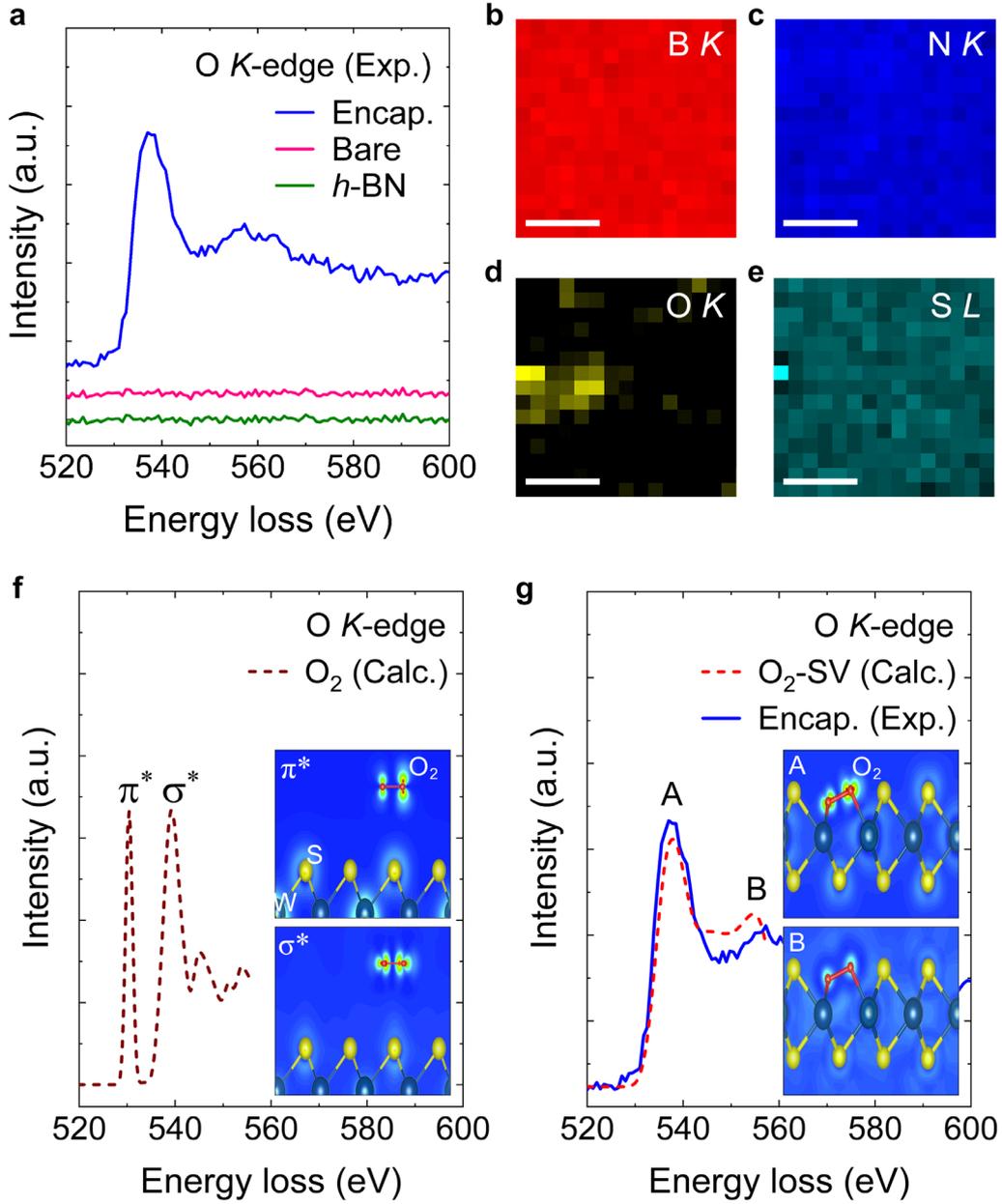

**Figure 2.** a) Experimental oxygen $K$-edge EELS spectra for the $h$-BN encapsulated $WS_2$, bare $WS_2$, and $h$-BN flake crystals. b-e) The EELS maps for the boron (b), nitrogen (c), oxygen (d) $K$-edge, and the sulfur (e) $L$-edge measured from $h$-BN encapsulated $WS_2$. The scale bars are 100 nm. f) The calculated oxygen $K$-edge EELS spectrum $\epsilon_2(\omega)$ for a physisorbed oxygen molecule on the pristine $WS_2$. g) The calculated oxygen $K$-edge EELS spectrum $\epsilon_2(\omega)$ (red dashed) for a chemisorbed oxygen molecule on the sulfur vacancy ($O_2$-SV) of the $WS_2$. The experimental oxygen $K$-edge spectrum (blue line) for the $h$-BN encapsulated $WS_2$ is in good agreement with the calculation result (red dashed). The insets of (f,g) represent the real-space orbital distributions $\rho(r)_E$ of the $\pi^*$ ($E \approx 530$ eV) and $\sigma^*$ ($E \approx 539$ eV) for (f) and the A ($E \approx 538$ eV) and B ($E \approx 555$ eV) for (g), respectively.

The investigation of the exciton recombination processes against the excitation power for the bare and *h*-BN encapsulated WS$_2$ crystals reveals that the oxygen fixation suppresses the nonradiative decay for the neutral excitons by stabilizing the free electron density of the WS$_2$. Figures 3a,b show the double-logarithmic plots of the neutral exciton (X$^0$) emission intensity of the bare and *h*-BN encapsulated WS$_2$ crystals as a function of the excitation power density under ambient air and vacuum conditions at room temperature, respectively. The bare WS$_2$ crystal in air exhibits a sublinear increase with an exponent of 0.71 in a power-law ($I \propto P^\alpha$, $P$: excitation power), while that in vacuum shows a smaller exponent of 0.32. The power dependence can be understood by introducing a simple rate equation model for the steady-state photoluminescence. The rate equation for exciton generation and recombination can be described by[23,24]

$$G = \frac{n_X}{\tau_X} + T n_X n_e + A n_X n_e + \gamma n_X^2 \qquad (1)$$

where $n_X$ is the neutral exciton density, $\tau_X$ is the exciton lifetime, $n_e$ is the electron density, $T$ is the trion formation coefficient, $A$ is the exciton-electron Auger coefficient, and $\gamma$ is the exciton–exciton annihilation coefficient. At our excitation power range, the exciton density is estimated to be from the low $10^{10}$ cm$^{-2}$ to the low $10^{11}$ cm$^{-2}$. In this range of exciton density, it has been known that the exciton-exciton annihilation process becomes negligible.[25] Furthermore, the recent work has shown that the photogenerated excitons are mostly converted to the trions in this range, indicating that the Auger process can also be neglected.[24] Thus, the predominant nonradiative decay channel of the neutral excitons is the exciton-to-trion conversion process (see Figure S13 in Supporting Information). When the free electron density increases with the power dependence of $P^\mu$, the neutral excitons are converted to the trions in proportional to $P^\mu$, and the neutral exciton emission intensity follows the power-law of $P^{1-\mu}$ (see Supporting Information S12). The sublinear increase in the exciton emission in bare WS$_2$ indicates that the free electron density increases with increasing excitation power density by releasing oxygen molecules chemisorbed on defect sites, promoting the trion conversion process. The smaller exponent further confirms the easier desorption under ambient vacuum condition. Interestingly, the *h*-BN encapsulated WS$_2$ shows a linear increase with an exponent of 0.99 for both the air and vacuum ambient conditions, as shown in Figure 3b. This indicates that *h*-BN encapsulation results in predominant neutral exciton emission over the decay channels without generating free electrons by desorption.[16,24] Figure 3c shows the photoluminescence intensity ratio (X$^-$/X$^0$) of the trion to the neutral exciton as a function of the excitation power density. The X$^-$/X$^0$ of bare WS$_2$ measured under ambient vacuum condition steeply increases from 0.38 to 4.18, while the $X^-/X^0$ under the air environment gradually increases from 0.25 to 0.50 over the excitation power density range from 0.03 to 0.66 kW/cm$^2$. This further evidences that the desorption process, which induces free electrons and exciton-to-trion conversion, is much facilitated in vacuum than in air. In striking contrast, the $X^-/X^0$ of $h$-BN encapsulated WS$_2$ is kept almost constant at approximately 0.02 over the excitation power densities under both the ambient air and vacuum conditions, which is greatly reduced by more than one order of magnitude compared to that of the bare WS$_2$. This clearly indicates that $h$-BN encapsulation prevents the desorption of oxygen molecules by the laser illumination, resulting in predominant recombination by neutral excitons. By considering the intensity weight ($I_{X^-}/I_{total}$) of the trion in the photoluminescence spectra, we estimated the free electron density as a function of the excitation power density by using the mass action law (see Supporting Information S14).[26,27] As shown in Figure 3d, the free electron density exhibits a very similar trend as $X^-/X^0$. For bare WS$_2$, the free electron density drastically increases and reaches $2.66 \times 10^{13}$ cm$^{-2}$ at an excitation power density of 0.66 kW/cm$^2$ in vacuum, which is close to the Mott density ($\sim 1 \times 10^{14}$ cm$^{-2}$),[28] while gradually increasing to $7.38 \times 10^{12}$ cm$^{-2}$ at an excitation power density of 4.99 kW/cm$^2$ in air. However, regardless of the ambient conditions, the free electron density of $h$-BN encapsulated WS$_2$ is maintained at $\sim 9 \times 10^{10}$ cm$^{-2}$ with increasing excitation power, resulting from the fixation of chemisorbed oxygen molecules by $h$-BN layers. Additonally, using the bare and $h$-BN encapsulated WS$_2$ capacitor devices, we estimated the free electron densities in the bare and $h$-BN encapsulated WS$_2$ at the vacuum and air environments. For the bare WS$_2$, the electron densities were estimated to be $1.15 \times 10^{13}$ cm$^{-2}$ (vacuum) and $3.69 \times 10^{12}$ cm$^{-2}$ (air) at $V_g \cong 0\ V$ (gate voltage), respectively, while those of $h$-BN encapsulated WS$_2$ were estimated to be $2.05 \times 10^{11}$ cm$^{-2}$ at both the vaccum and air environments. These electron densities were in good agreement with those estimated using the mass-action law. From the difference in free electron densities in-between the vacuum and air environments, we estimated the number of desorbed/adsorbed oxygen molecules on the sulfur vacanies to be $7.17 \times 10^{12}$ cm$^{-2}$ (see Supporting Information S15).

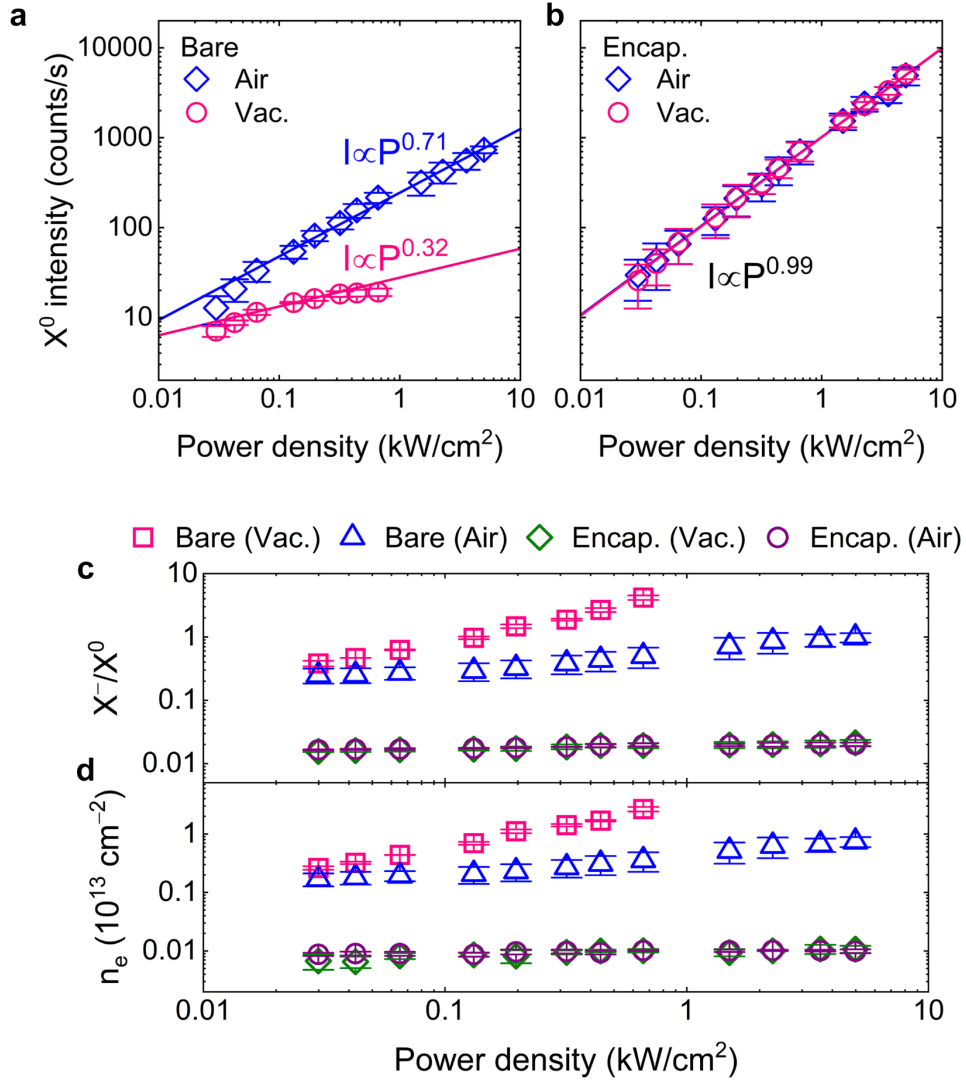

**Figure 3.** a,b) Neutral exciton ($X^0$) emission intensity of the bare (a) and *h*-BN encapsulated (b) WS$_2$ crystals as a function of the excitation power density at room temperature. The blue diamond and pink circle symbols indicate the power-dependent exciton emission intensity measured under ambient air and vacuum conditions. c) Photoluminescence intensity ratio ($X^-/X^0$) of the trion to the neutral exciton as a function of the excitation power density. d) Free electron density as a function of the excitation power density by using the mass action law. The pink square (blue triangle) and olive diamond (purple circle) symbols represent the $X^-/X^0$ and free electron density of the bare and *h*-BN encapsulated WS$_2$ under vacuum (air) ambient conditions, respectively. The error bars in (a−d) indicate the standard deviation of the measured data.

The nonradiative decay channel of neutral excitons can be attributed to exciton-to-trion conversion at our excitation power range. From the rate equation (1), the total recombination rate ($R_T$) can be described by $R_T = R_0 + R_A n_X$, where $R_0$ is the density-independent recombination rate and $R_A$ is the exciton annihilation rate constant due to the exciton-to-trion conversion process. The role of the oxygen fixation can be quantitatively understood by estimating the exciton annihilation rate constant ($R_A$). To measure the exciton lifetime with increasing excitation power, we carried out time-resolved photoluminescence spectroscopy. For bare WS$_2$, the photoluminescence decay curves of neutral excitons under air ambient condition show a steep shortening of the decay time from 74 ps to 43 ps with increasing pump fluence from 27 to 959 nJ/cm$^2$, as shown in Figure 4a. However, the exciton decay curves for $h$-BN encapsulated WS$_2$ show a slight decrease in the lifetime above the pump fluence of 421 nJ/cm$^2$, as shown in Figure 4b. It is noteworthy that the exciton lifetime in $h$-BN encapsulated WS$_2$ becomes longer than that of bare WS$_2$, showing exciton lifetime of 136 ps in air at a pump fluence of 27 nJ/cm$^2$. The decay spectra at vacuum are shown in Figure S16 in Supporting Information. The increase in exciton lifetime in the $h$-BN encapsulated WS$_2$ strongly suggests that the nonradiative decay by the trion formation, which occurs on a very fast time scale of a few ps, is significantly inhibited due to the passivation of defects by the oxygen fixation.[23,29,30] To quantitatively evaluate the exciton annihilation rate constant ($R_A$), the exciton density-induced recombination rate ($\tau^{-1}$) was plotted as a function of the exciton density using the measured exciton lifetimes (see Supporting Information S17).[30] Note that the exciton density was estimated by calculating the net absorption in the monolayer WS$_2$ for the pump fluence.[31,32] As presented in Figures 4c,d, the $h$-BN encapsulated WS$_2$ crystals exhibit an exciton annihilation rate constant of 8.3 × 10$^{-3}$ cm$^2$s$^{-1}$ (7.8 × 10$^{-3}$ cm$^2$s$^{-1}$), while the bare WS$_2$ crystals show a value of 8.0 × 10$^{-2}$ cm$^2$s$^{-1}$ (3.2 × 10$^{-1}$ cm$^2$s$^{-1}$) under air (vacuum) ambient conditions. As expected, the exciton annihilation rate constant of the $h$-BN encapsulated WS$_2$ is remarkably reduced by approximately two orders of magnitude compared to that of the bare WS$_2$. These results are attributed to the suppression of exciton-to-trion conversion process in the $h$-BN encapsulated WS$_2$ due to the greatly lowered and stabilized free electron density. This fact can be additionally verified by investigating the decay dynamics of neutral excitons with an electrostatic doping in the $h$-BN encapsulated WS$_2$ capacitor devices at a fixed excitation power. Figure 4e shows the gate-voltage-dependent photoluminescence spectral map in the $h$-BN encapsulated WS$_2$ capacitor devices, showing that the charge neutral point is determined at $V_g \cong 0\ V$. The increase in the gate voltage ($V_g > 0\ V$) gives rise to the gradual decrease in the emission intensity for the neutral exciton and, simultaneously, the increase in the emission intensity for the trion. This indicates that the increase in free electron density facilitates the exciton-to-trion conversion process, leading to the nonradiative decay of neutral excitons. The gate-voltage-dependent photoluminescence decay curves of neutral excitons also clearly show that the increase in the free electron density promotes the exciton annihiliation. As shown in Figure 4f, the decay time of neutral excitons in the *h*-BN encapsulated WS$_2$ capacitor devices becomes steeply shorten from 139 ps to 36 ps for increasing the gate voltage from 0.5 V to 0.9 V. The exciton annihilation rate constant with the electrostatic doping is estimated to be $1.3 \times 10^{-1}$ cm$^2$s$^{-1}$ (see Supporting Information S18), which is similar to that of bare WS$_2$ under the vacuum ambient condition.

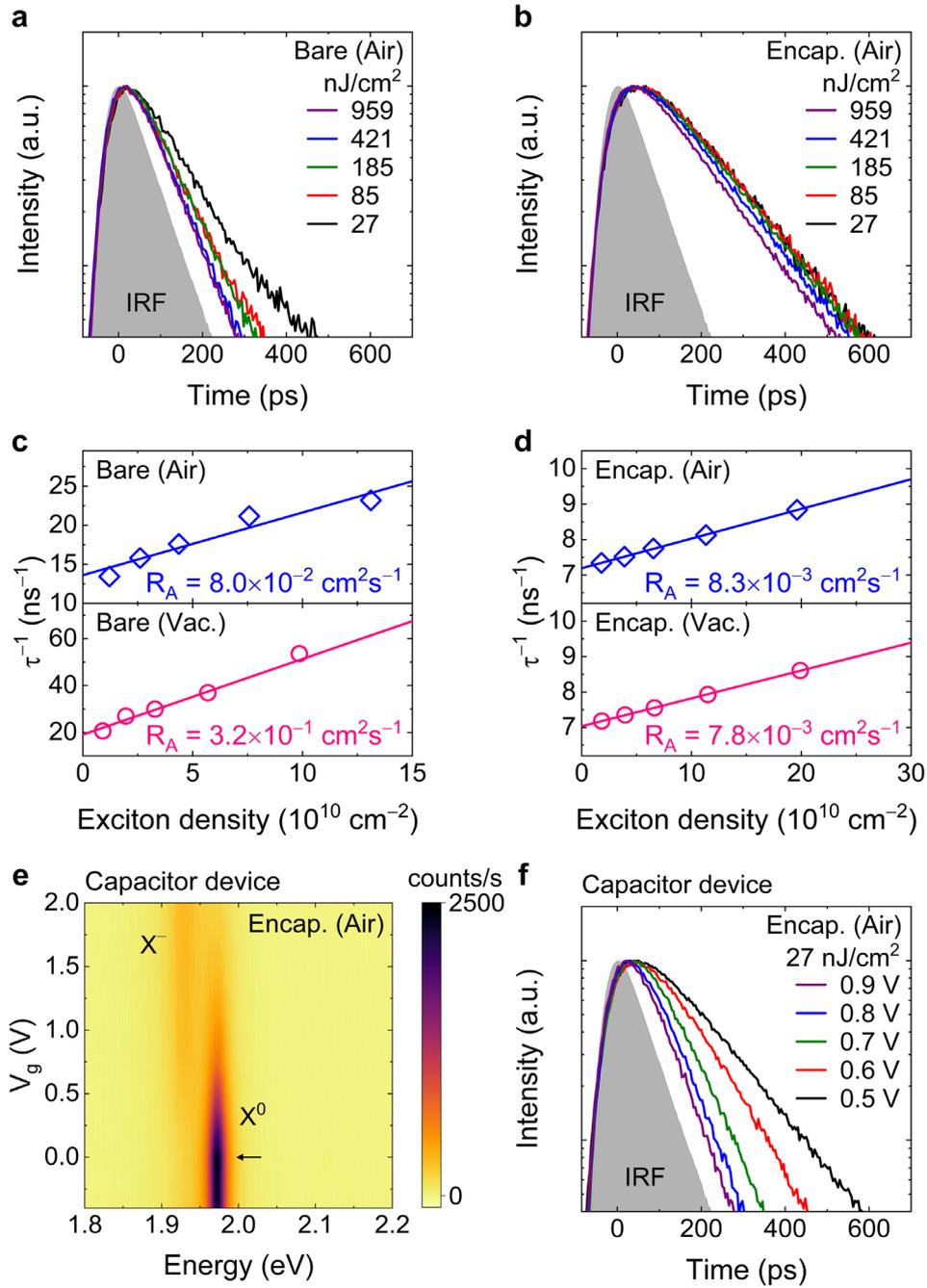

**Figure 4.** a,b) Photoluminescence decay curves of the neutral excitons measured as a function of the energy fluence for bare (a) and $h$-BN encapsulated (b) WS$_2$ under ambient air condition. c,d) Exciton density-induced recombination rate ($\tau^{-1}$) for the bare (c) and $h$-BN encapsulated (d) WS$_2$ under ambient air and vacuum conditions, resulting in the exciton annihilation rate constant ($R_A$) due to exciton-to-trion conversion process by a linear fit. e) Gate-voltage-dependent photoluminescence spectral map of the $h$-BN encapsulated WS$_2$ capacitor devices. The black arrow indicates the charge neutral point of the $h$-BN encapsulated WS$_2$ capacitor device. f) Gate-voltage-dependent photoluminescence decay curves of the neutral excitons in the $h$-BN encapsulated WS$_2$ capacitor device.

The *h*-BN encapsulation gives rise to an almost constant level of the free electron density in WS$_2$ crystals for elevated excitation powers, which can provide a stable and robust valley polarization against various excitation conditions. In contrast, for bare WS$_2$, the drastic increase in free electron density with increasing excitation would cause a large variation in the valley polarization due to the change in decay dynamics of the neutral excitons caused by the exciton-to-trion conversion.[33,34] To investigate the effect of *h*-BN encapsulation on the valley polarization, we carried out circular polarization-resolved photoluminescence measurements as a function of the excitation power density at 77 K. Figure 5a show the circularly polarized photoluminescence spectra measured from the bare (top panel) and *h*-BN encapsulated (bottom panel) WS$_2$, respectively. As a result, the degree of valley polarization with increasing excitation power shows a very different trend for the bare and *h*-BN encapsulated WS$_2$ crystals, as shown in Figure 5b. An important distinction is that the *h*-BN encapsulated WS$_2$ exhibits a stable valley polarization ratio at a constant level, whereas that of the bare WS$_2$ shows a large variation by changing the excitation power. The valley polarization can be described by the following equation $P_V = P_0/(1 + 2\tau_X/\tau_V)$,[35] where $P_0$ is the initial valley polarization, $\tau_X$ is the valley exciton lifetime, and $\tau_V$ is the valley relaxation time. The initial valley polarization ($P_0$) given by the optical selection rules can be assumed to be unity,[5] meaning that the valley polarization ($P_V$) is then mainly governed by the competition between the exciton lifetime and the valley relaxation time.[35] As shown in Figure 5b, the degree of valley polarization of bare WS$_2$ is 1.5 times higher than that of *h*-BN encapsulated WS$_2$ at the lowest excitation power density of 0.17 kW/cm$^2$. This can be attributed to the fact that in bare WS$_2$, the neutral excitons decay rapidly within the valley, rather than an intervalley scattering, due to a shortened exciton lifetime caused by a higher exciton-to-trion conversion rate. In addition, the large variation in the valley polarization of bare WS$_2$ can be understood as a result of the change in the exciton lifetime by the accelerated trion conversion process with increasing excitation power, as shown in Figures 4a and c.[33,34] Accordingly, the valley polarization can be increased in the bare WS$_2$ for elevated excitation powers. For the *h*-BN encapsulated WS$_2$, however, the exciton-to-trion conversion and the exciton lifetime are maintained at almost constant levels, resulting in a stable valley polarization ratio for elevated excitation powers. The large variation in the valley polarization is also observed in the *h*-BN encapsulated WS$_2$ capacitor devices with the electrostatic doping (Figure 5c), showing the drastic decrease in the exciton lifetime as the gate voltage increases (see Supporting Information S19). This also confirms that the change in the free electron density leads to the large variation in the valley polarization.

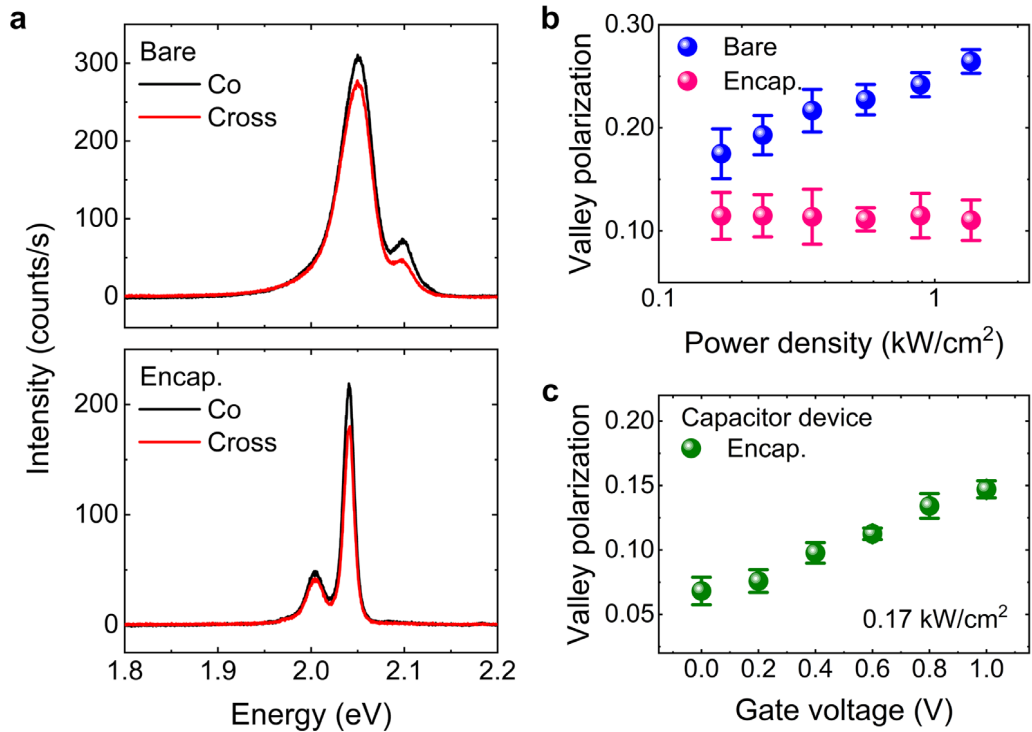

**Figure 5.** a) Circularly polarized photoluminescence spectra for the bare (top panel) and *h*-BN encapsulated (bottom panel) WS$_2$ measured at 77 K. b) Degree of valley polarization for the exciton emission in the bare and *h*-BN encapsulated WS$_2$ taken as a function of the excitation power density. c) Degree of valley polarization for the neutral excitons in the *h*-BN encapsulated WS$_2$ capacitor devices taken as a function of the gate voltage. The error bars marked in (b,c) exhibit the standard deviation of the measured valley polarization values.

## 3. Conclusion

In conclusion, we have demonstrated that *h*-BN encapsulation greatly removes the defect-related gap states by stabilizing the chemisorbed oxygen molecules onto the defects of monolayer $WS_2$ crystals, that are provided during the interactions between $WS_2$ and atmosphere. It is clearly shown that the oxygen molecules are chemisorbed onto the defects of $WS_2$ crystals and are fixated by *h*-BN encapsulation with excluding a possibility of oxygen molecules trapped in bubbles or wrinkles formed at the interface between $WS_2$ and *h*-BN, as confirmed by the EELS study. Optical spectroscopic studies show that *h*-BN encapsulation prevents the desorption of oxygen molecules over various excitation and ambient conditions, resulting in a greatly lowered and stabilized free electron density in monolayer $WS_2$ crystals. This suppresses the exciton annihilation processes by two orders of magnitude compared to that of bare $WS_2$. Furthermore, due to the stabilized free electron density in the *h*-BN encapsulated $WS_2$ crystals, the valley polarization becomes robust against the various excitation and ambient conditions. Our findings provide insight into the role of *h*-BN encapsulation and open up the possibility to control the defect states in 2D semiconductors through adsorbate-engineered 2D heterostructures.

## 4. Experimental Section

*Sample preparation:* The monolayer $WS_2$ crystals were grown on a sapphire substrate by chemical vapor deposition methods,[19] and the *h*-BN flakes with a thickness of ~40 nm were prepared by mechanical exfoliations from bulk *h*-BN single crystals. The *h*-BN encapsulated $WS_2$ structure was fabricated by sequential pick-up processes using the dry van der Waals stacking method. Elvacite resin or polycarbonate (PC) was utilized as a polymer stamp for the pick-up of layered materials. The assembled *h*-BN/$WS_2$/*h*-BN structures and picked-up $WS_2$ crystals were released onto the line trenches by melting the polymer stamp, where the line trenches were fabricated through conventional photolithography and reactive ion etching processes using 400-nm-thick $SiO_2$-coated Si substrates. The polymer stamps were removed by immersing the fabricated sample in chloroform. Finally, both the *h*-BN encapsulated and the bare $WS_2$ onto line trenches were annealed at 350 °C to improve the coupling between the stacked layers and remove transfer residues.

*Optical measurements:* The steady-state and time-resolved photoluminescence measurements were performed using a home-built confocal microphotoluminescence system. Using a 40× (0.6 NA) objective (Nikon), the excitation beam was focused, and the signal from the samples was collected through an optical fiber on the focal image plane. For the steady-state photoluminescence measurements, an argon-ion laser with a wavelength of 457.9 nm (continuous wave) was used as an excitation source. The photoluminescence spectra were resolved by a spectrometer (Acton SpectraPro 500i with 0.5 m focal length and 1200 grooves/mm grating) equipped with a charge-coupled device (CCD) detector (Princeton Instruments, 512 × 2048 pixels). For the circular polarization measurements, a combination of linear polarizer and quarter waveplate was used to generate circularly polarized excitation light, while another pair of linear polarizer and quarter waveplate was set for a polarization analyzer before collecting the signal through the slit of the spectrometer. The degree of valley polarization is defined as $\rho = [PL(\sigma^+) - PL(\sigma^-)]/[PL(\sigma^+) + PL(\sigma^-)]$, where $PL(\sigma^\pm)$ are the photoluminescence intensity for $\sigma^+$ and $\sigma^-$ polarized light components under excitation with a $\sigma^+$ or $\sigma^-$ polarized laser beam. The time-resolved photoluminescence measurements were carried out using a picosecond pulsed diode laser (PicoQuant, LDH-P-FA-355) with a wavelength of 355 nm (FWHM = 56 ps) and repetition rate of 40 MHz. The exciton lifetimes were measured using a hybrid photomultiplier detector (PicoQuant, PMA hybrid series) and a time-correlated single photon counting system (PicoQuant).

*Scanning transmission electron microscopy (STEM) and EELS analysis*: STEM was used by Monochromated ARM-200F (NEO-ARM) in Korea Basic Science Institute (KBSI) operated at 200 kV. Gatan imaging filter (GIF) Continuum HR-1066 spectrometer was used to collect electron energy loss spectra.

**Acknowledgements**
This work was supported by the Basic Science Research Program (2019R1A2C1088525) and the BrainLink program (RS-2023-00236798) through the National Research Foundation of Korea, by the DGIST R&D Program (23-CoE-NT-01 and 23-HRHR+-03) funded by the Ministry of Science and ICT of the Korean Government. K.W. and T.T. acknowledge support from JSPS KAKENHI (Grant Numbers 19H05790, 20H00354 and 21H05233) and A3 Foresight by JSPS. Y.K. was supported by National Research Foundation of Korea (2020R1C1C1006914). J.H.J. and H.S.C. were supported by the Technology Innovation Program (20010542) funded by the MOTIE, Korea. The authors thank the computational support from the Center for Advanced Compuation (CAC) at Korea Insitute for Advanced Study (KIAS). Y.K. supported by a KIAS individual Grant (PG088601).

# Supporting Information

**Defect Passivation of 2D Semiconductors by Fixating Chemisorbed Oxygen Molecules via *h*-BN Encapsulations**

*Jin-Woo Jung, Hyeon-Seo Choi, Young-Jun Lee, Youngjae Kim, Takashi Taniguchi, Kenji Watanabe, Min-Yeong Choi, Jae Hyuck Jang, Hee-Suk Chung, Dohun Kim, Youngwook Kim, and Chang-Hee Cho\**

Jin-Woo Jung, Hyeon-Seo Choi, Young-Jun Lee, Dohun Kim, Youngwook Kim, Chang-Hee Cho
Department of Physics and Chemistry, Daegu Gyeongbuk Institute of Science and Technology (DGIST), Daegu 42988, South Korea
E-mail: chcho@dgist.ac.kr

Youngjae Kim
School of Physics, Korea Institute for Advanced Study (KIAS), Seoul 02455, South Korea

Takashi Taniguchi
International Center for Materials Nanoarchitectonics, National Institute for Materials Science, Tsukuba 305-0044, Japan

Kenji Watanabe
Research Center for Functional Materials, National Institute for Materials Science, Tsukuba 305-0044, Japan

Min-Yeong Choi, Jae Hyuck Jang, Hee-Suk Chung
Electron Microscopy and Spectroscopy Team, Korea Basic Science Institute, Daejeon 34133, South Korea

Jae Hyuck Jang
Graduate School of Analytic Science and Technology, Chungnam National University, Daejeon 34134, South Korea

**S1. Kinetic barriers for $O_2$-chemisorption and $O_2$-dissociative chemisorption processes**

We performed density functional theory calculations implemented in the quantum espresso code, employing Optimized Norm-Conserving Vanderbilt (ONCV) pseudopotentials and the nudged elastic band (NEB) methods for 3 by 3 superstructures of monolayer $WS_2$. These calculations, based on the grimme-D2 van der Waals corrections, reveal the relative energy evolution for each reaction path from the initial configurations to the final configurations.

Figures S1a and S1b show the kinetic barriers for the $O_2$-chemisorption and $O_2$-dissociative chemisorption processes in the monolayer $WS_2$ (a) and $WTe_2$ (b) with the sulfur vacancy (SV). In the case of the $WS_2$, the kinetic barrier for the $O_2$-chemisorption (0.56 eV) is lower than that for the $O_2$-dissociative chemisorption process (0.76 eV), as shown in Figure S1a. From the transition state theory, the reaction rate is given by $k \cong f\exp(-E_b/k_B T)$, where $k$ is the reaction rate, $f$ is the attempt frequency, $E_b$ is the barrier, $k_B$ is the Bolzmann constant, and $T$ is the temperature. The attempt frequency can be approximated by the value of $10^{12}$ s$^{-1}$.[1] The reaction rate ($T$ = 300 K) for the $O_2$-chemisorption and dissociative chemisorption processes is estimated to be approximately 180 s$^{-1}$ ($E_b$ = 0.56 eV) and 0.17 s$^{-1}$ ($E_b$ = 0.76 eV), respectively. This result indicates that the probability of the $O_2$-chemisorption is 1000 times higher than that of the $O_2$-dissociative chemisorption. Thus, the oxygen chemisorption on the monolayer $WS_2$ crystals would have the final configuration of the $O_2$-chemisorption rather than the $O_2$-dissociative chemisorption.[2]

On the other hand, in the case of the $WTe_2$ (Figure S1b), there are no kinetic barriers (0.00 eV) for both the $O_2$-chemisorption and $O_2$-dissociative chemisorption processes, indicating that the oxygen chemisorption spontaneously occurs toward the $O_2$-dissociative chemisorption process in the case of $WTe_2$. These results are in good agreement with previous theoretical studies, showing that the type of the oxygen chemisorption depends on TMD materials.[2]

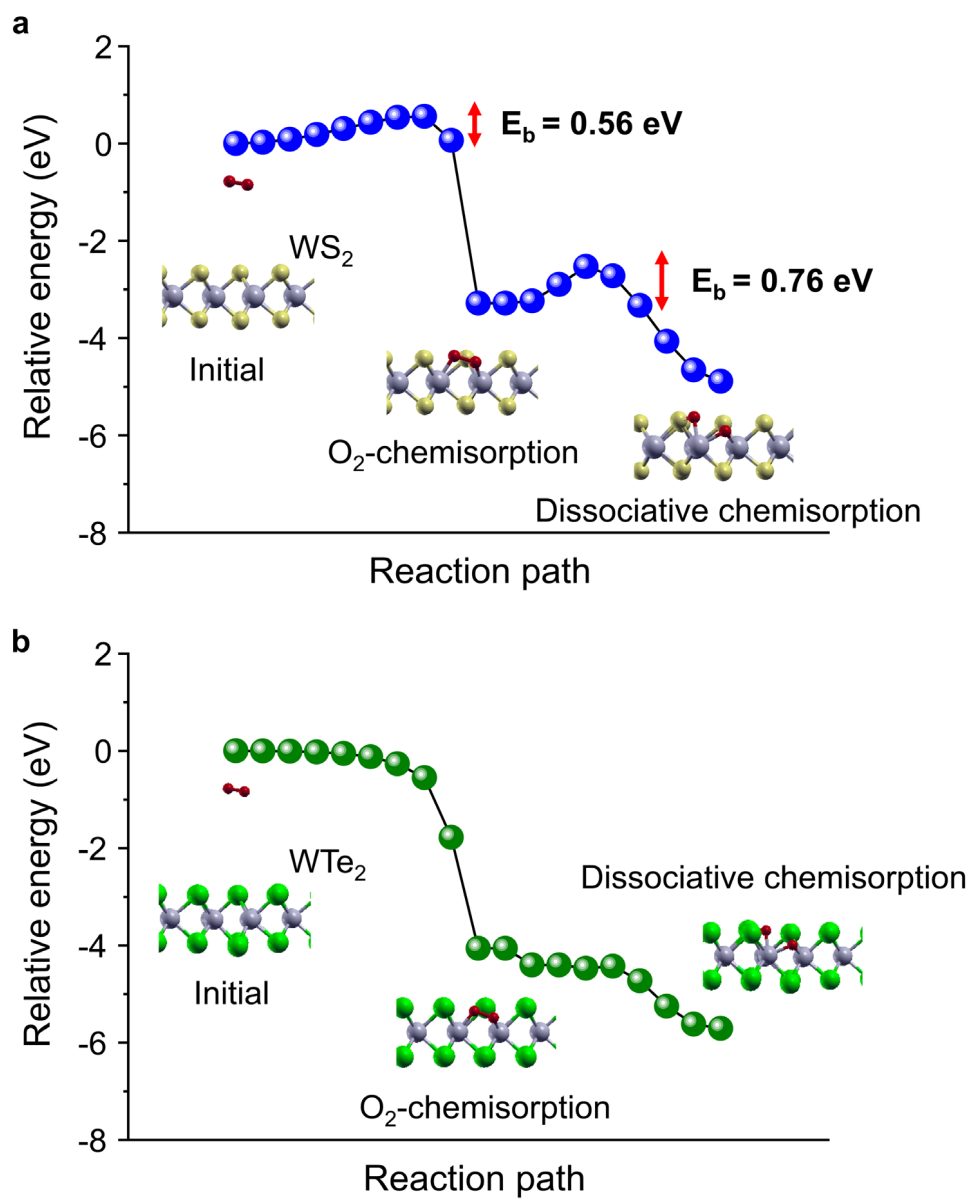

**Figure S1.** (a,b) Calculated reaction path and kinetic barriers for the $O_2$-chemisorption and $O_2$-dissociative chemisorption processes in the monolayer $WS_2$ (a) and $WTe_2$ (b) with the sulfur vacancy, respectively.

**S2 and S3. Molecular interactions with the pristine surface and sulfur vacancy of WS$_2$**

Figures S2 and S3 reveal the relative energy evolutions as a function of the reaction path for N$_2$, O$_2$, and H$_2$O with both the pristine WS$_2$ and SV of WS$_2$. In this reaction path, the decrease in relative energy indicates that the molecular adsorption proceeds toward a physisorption, corresponding to an exothermic process.[3] Meanwhile, an increase in relative energy exhibits that the molecular adsorption proceeds toward a chemisorption, which leads to an activation barrier in the chemical bonding sequence.

As shown in Figure S2, the N$_2$ molecule does not achieve a structurally favorable state when chemisorbed onto the SV. However, the N$_2$ exhibits a preference for physisorption (−0.08 eV) just prior to forming a chemical bond with surrounding tungsten atoms. Conversely, the O$_2$ starting from a weakly physisorbed state displays a minor repulsive barrier of ~0.56 eV just before chemically bonding. Subsequently, it attains a fully stable chemisorption state with the SV site, resulting a favorable energy of −3.28 eV. On the other hand, the H$_2$O prefers a physisorption-like configuration with the SV of WS$_2$, which does not show a chemisorption configuration.

Figure S3 shows the first-principles NEB calculation results for the molecular interaction with pristine WS$_2$. In case of the physisorption, all three molecules can be weakly physisorbed with the pristine WS$_2$. However, the weak adsorption energies in the final states lead to unstable physisorption configurations, resulting in the easy desorption of physisorbed molecules on pristine WS$_2$ surface. Thus the physisorption has virtually no influence on the electronic and optical properties of the WS$_2$ monolayer.

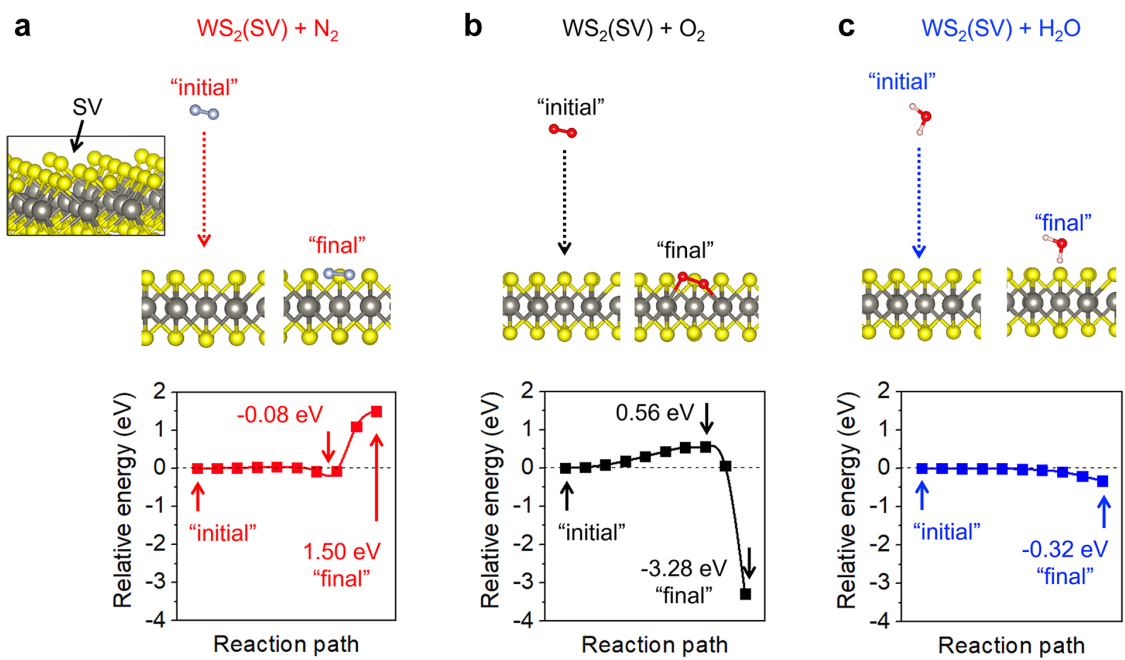

**Figure S2.** First-principles NEB calculations for molecular interactions with SV of $WS_2$. (a−c) The relative energy evolution of the $N_2$ (a), $O_2$ (b), and $H_2O$ (c) for each reaction path from the initial configurations to the final configurations. The relative energy is defined as the total energy difference compared to that of the initial. Adsorption of a molecule at given surface has two different forms, physisorption and chemisorption. The physisorption reflects an exothermic process, resulting in a decrease in relative energy. The chemisorption, on the other hand, exhibits an activation barrier in the chemical bonding sequence, leading to an increase in relative energy. (a) The $N_2$ molecule does not achieve a structurally favorable chemisorption state when adsorbed onto the SV. (b) The $O_2$ molecule can be only chemisorbed onto the SV and attain a fully stable chemisorption state (−3.28 eV). (c) The $H_2O$ molecule prefers a physisorption-like configuration with the SV of $WS_2$, which does not show a chemisorption.

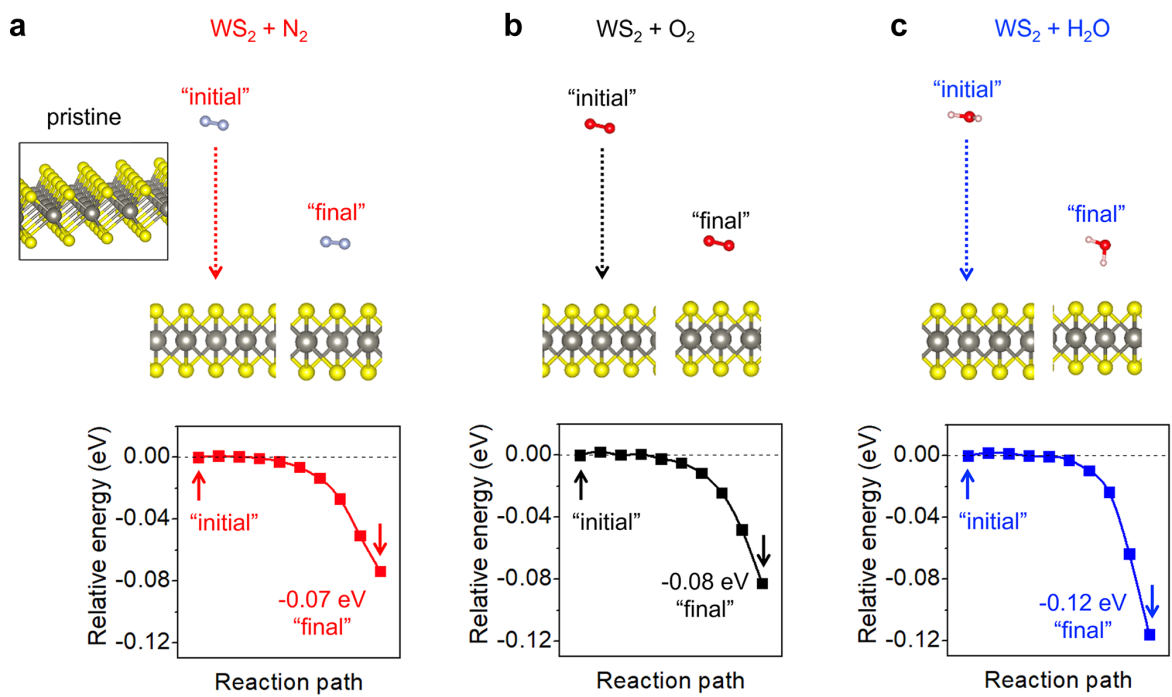

**Figure S3.** First-principles NEB calculations for molecular interactions with pristine WS$_2$. (a–c) The relative energy evolution of the N$_2$ (a), O$_2$ (b), and H$_2$O (c) for each reaction path from the initial configurations to the final configurations.

## S4. Bare and *h*-BN encapsulated WS$_2$ suspended on the line trenches

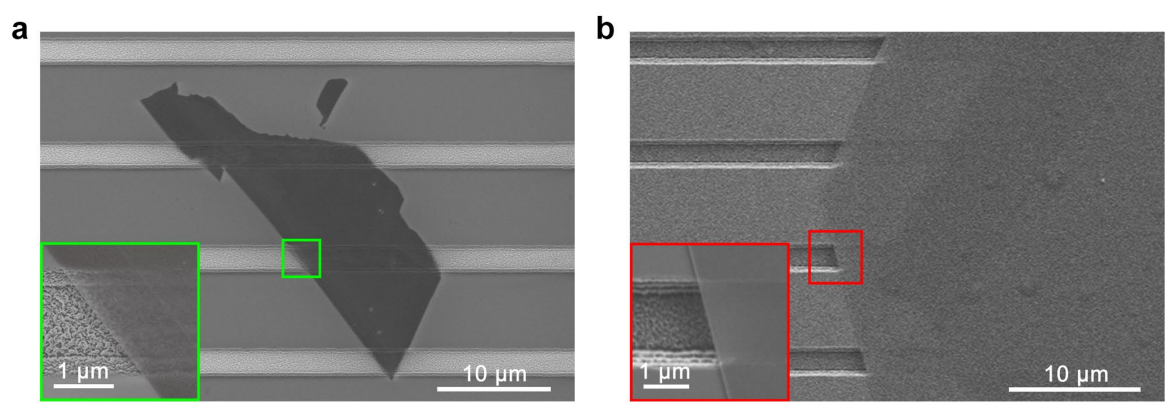

**Figure S4.** a,b) Scanning electron microscope images of the bare and *h*-BN encapsulated WS$_2$ crystals on the line trenches. The insets of (a) and (b) are the zoom-in images for the square marks of green and red boxes.

## S5. Exciton species in the bare and *h*-BN encapsulated WS$_2$

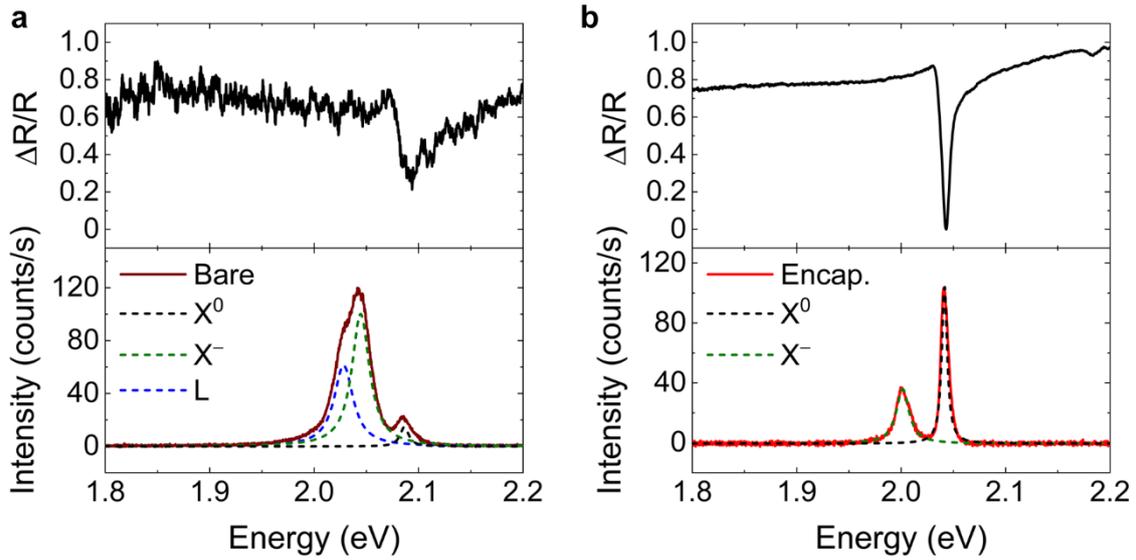

**Figure S5.** a,b) Differential reflectance and photoluminescence spectra measured from the bare (a) and *h*-BN encapsulated (b) WS$_2$ at the cryogenic temperature of 77 K. From the differential reflectance measurements, the exciton energy for the bare and *h*-BN encapsulated WS$_2$ was assigned to be 2.087 and 2.042 eV, respectively. By considering the peak separation between the exciton species based on the literatures,[4,5] the photoluminescence spectrum for the bare WS$_2$ is deconvoluted to three excitonic species corresponding to the neutral exciton (X$^0$), trion (X$^-$), and defect-related trapped exciton (L) states at 2.087, 2.042, and 2.026 eV, respectively. For the *h*-BN encapsulated WS$_2$, the two exciton species are assigned to be the neutral exciton (X$^0$), trion (X$^-$) at 2.042 and 2.001 eV, respectively. Note that the redshift of the exciton and trion energy in the *h*-BN encapsulated WS$_2$ compared with the bare WS$_2$ is due to the increase in the dielectric constant of the environment by the *h*-BN encapsulation.[6]

## S6. Excitonic spectra of the bare and *h*-BN encapsulated WS$_2$ under the different ambient conditions

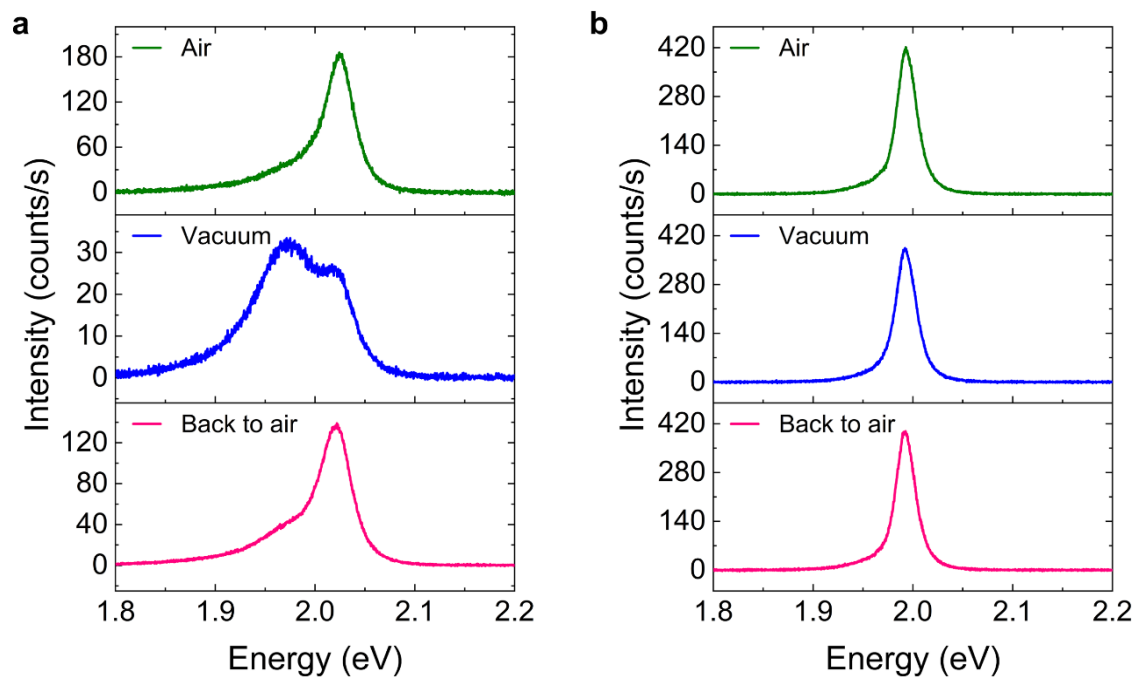

**Figure S6.** a,b) Photoluminescence spectra for the bare (a) and *h*-BN encapsulated (b) WS$_2$ measured under the ambient conditions of air, vacuum, and then air again. The photoluminescence spectral features of bare WS$_2$ are significantly altered according to the change in ambient condition, while those of *h*-BN encapsulated WS$_2$ are kept almost constant regardless of the change in ambient condition.

**S7. *h*-BN encapsulated WS₂ fabricated under an inert gas environment**

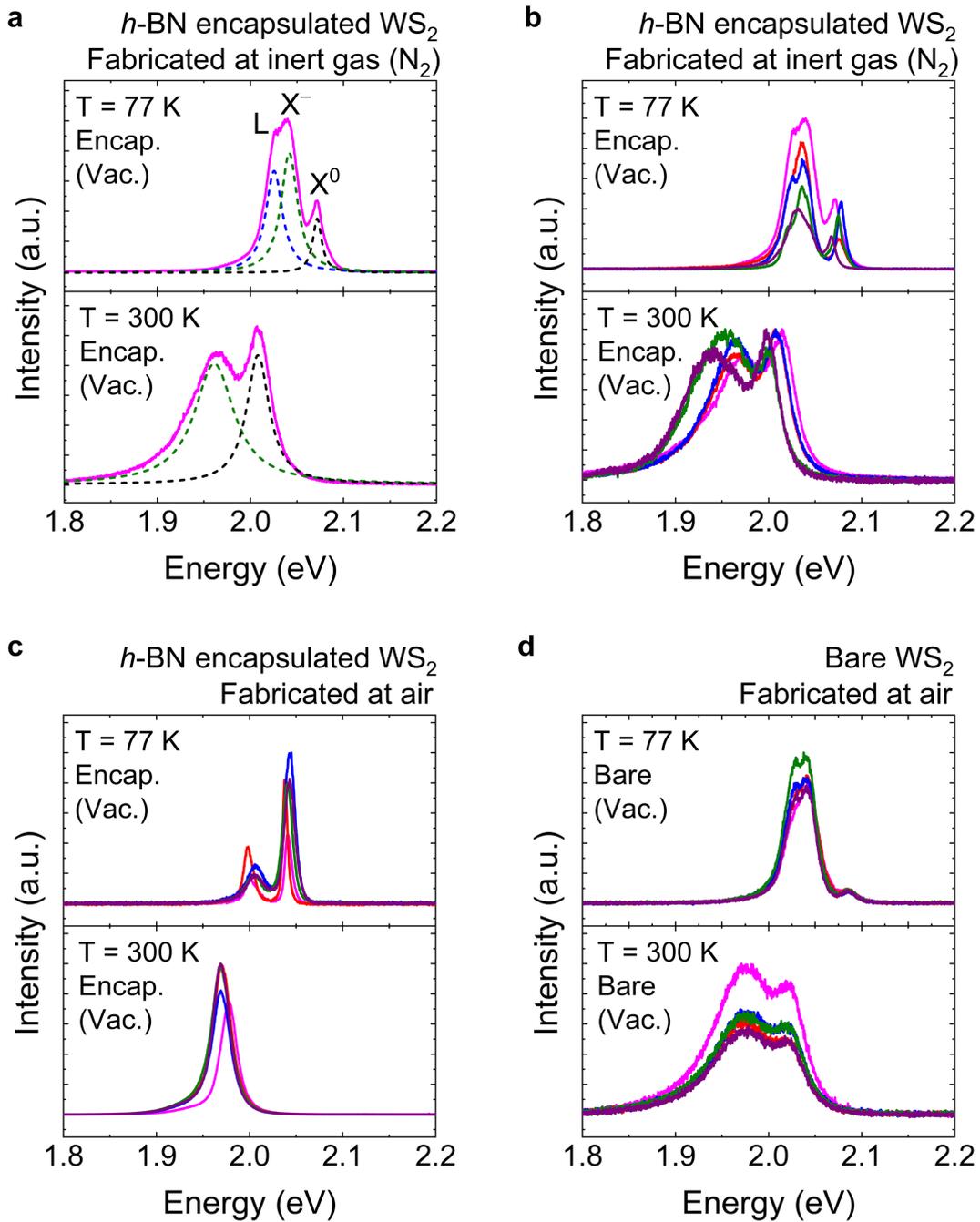

**Figure S7.** a) Photoluminescence spectra measured at the temperature of 77 K (top panel) and 300 K (bottom panel) for the *h*-BN encapsulated WS₂ fabricated under an inert environment. The black, olive, and blue dashed lines represent the neutral exciton ($X^0$), trion ($X^-$), and defect-related trapped exciton (L) states, respectively. b-d) Five representative spectra measured over many *h*-BN encapsulated WS₂ samples fabricated under the inert (b) and air (c) environments as well as the bare WS₂ (d) fabricated under the air environment.

## S8. Spectral change in the exfoliated monolayer WS$_2$ and WSe$_2$ with a lower density of chalcogen vacancies by varying the ambient conditions

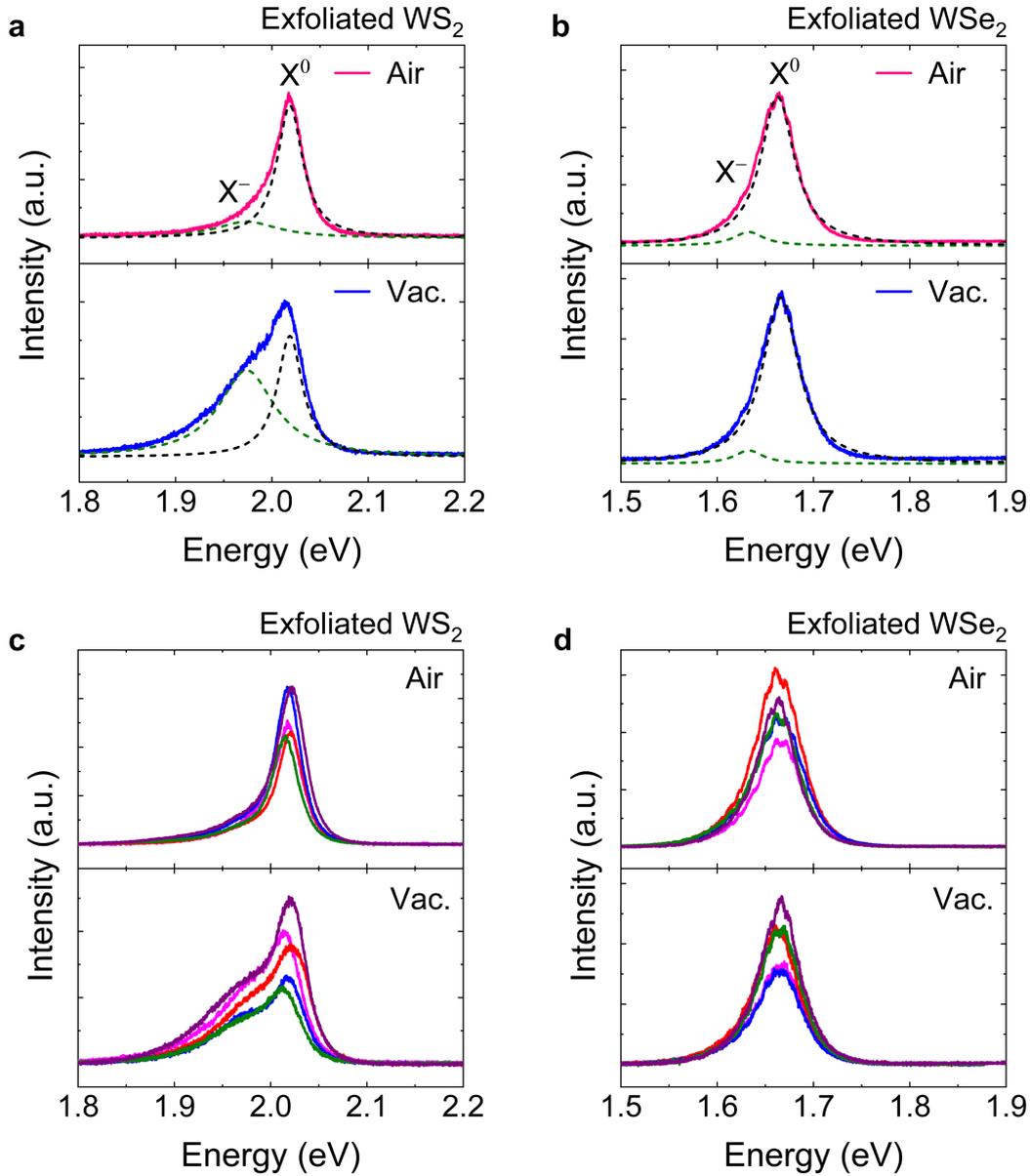

**Figure S8.** a,b) Photoluminescence spectra measured under the air (top panel) and vacuum (bottom panel) ambient conditions for the exfoliated monolayer WS$_2$ (a) and WSe$_2$ (b). The black and olive dashed lines represent the neutral exciton ($X^0$) and trion ($X^-$), respectively. c,d) Five representative spectra showing almost the same features over many WS$_2$ (c) and WSe$_2$ (d) samples under the variation of the air (top panel) and vacuum (bottom panel) ambient conditions.

## S9. Electron energy loss spectroscopy spectra

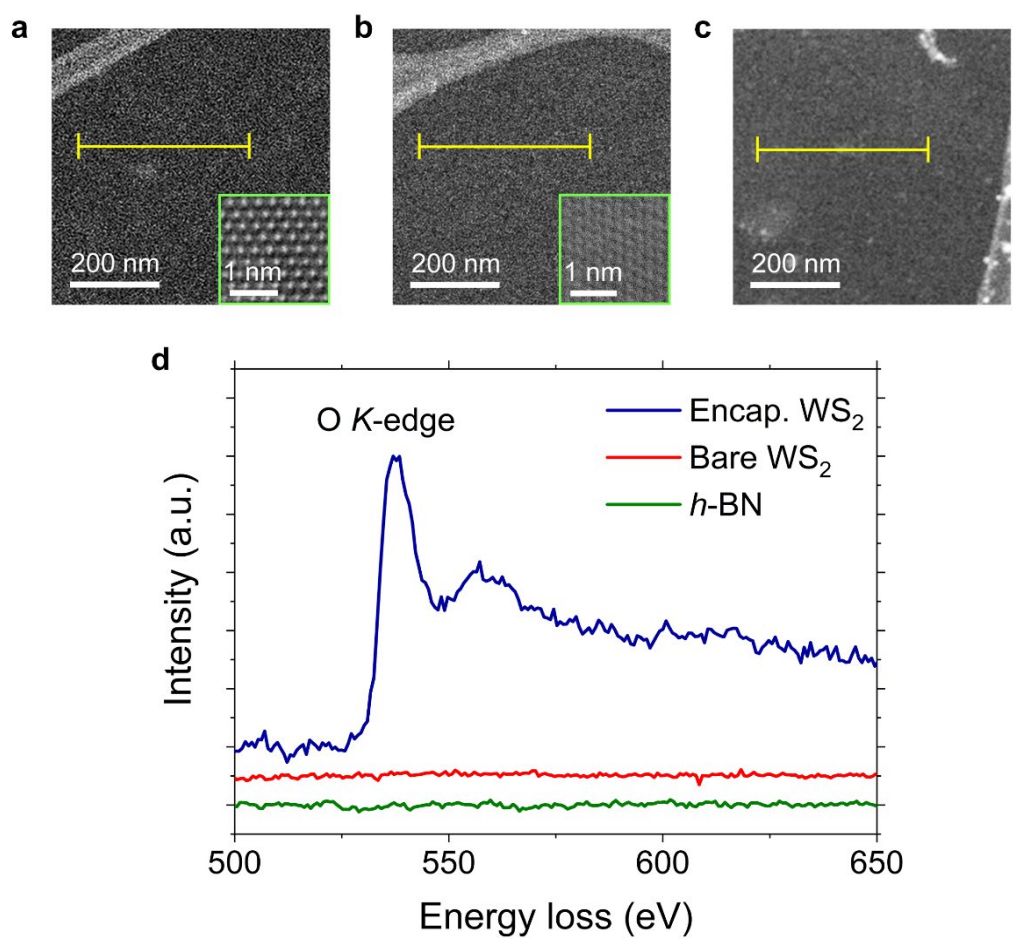

**Figure S9.** a–c) Annular dark field (ADF)-scanning transmission electron microscopy (STEM) images for bare WS$_2$ (a), *h*-BN flake (b), and *h*-BN encapsulated WS$_2$ crystals (c). Insets of (a,b) present the atomic-resolution ADF-STEM images for the bare WS$_2$ and *h*-BN flake. d) Oxygen *K*-edge electron energy loss spectroscopy (EELS) spectra for the *h*-BN encapsulated WS$_2$, bare WS$_2$, and *h*-BN flake. The EELS spectra are obtained from line-scanning for the regions (0.4 μm) marked as yellow bars in (a–c), and are acquired for a dwell time of 0.04 s per pixel. Note that in the case of the *h*-BN encapsulated WS$_2$, the strong boron and nitrogen signals in a few-nanometer-thick top and bottom *h*-BN layers encapsulating monolayer WS$_2$ interfere with the ability to obtain clear atomic-resolution STEM images of the monolayer WS$_2$. This makes it very challenging to clearly visualize the WS$_2$ plane in STEM images for *h*-BN encapsulated WS$_2$. More importantly, directly observing light atoms such as oxygen is challenging due to their easy knockout under the electron beam irradiation as well as the very weak TEM contrast, as widely reported in the previous works.[7,8]

**S10. Electron energy loss spectroscopy maps for *h*-BN encapsulated WS₂**

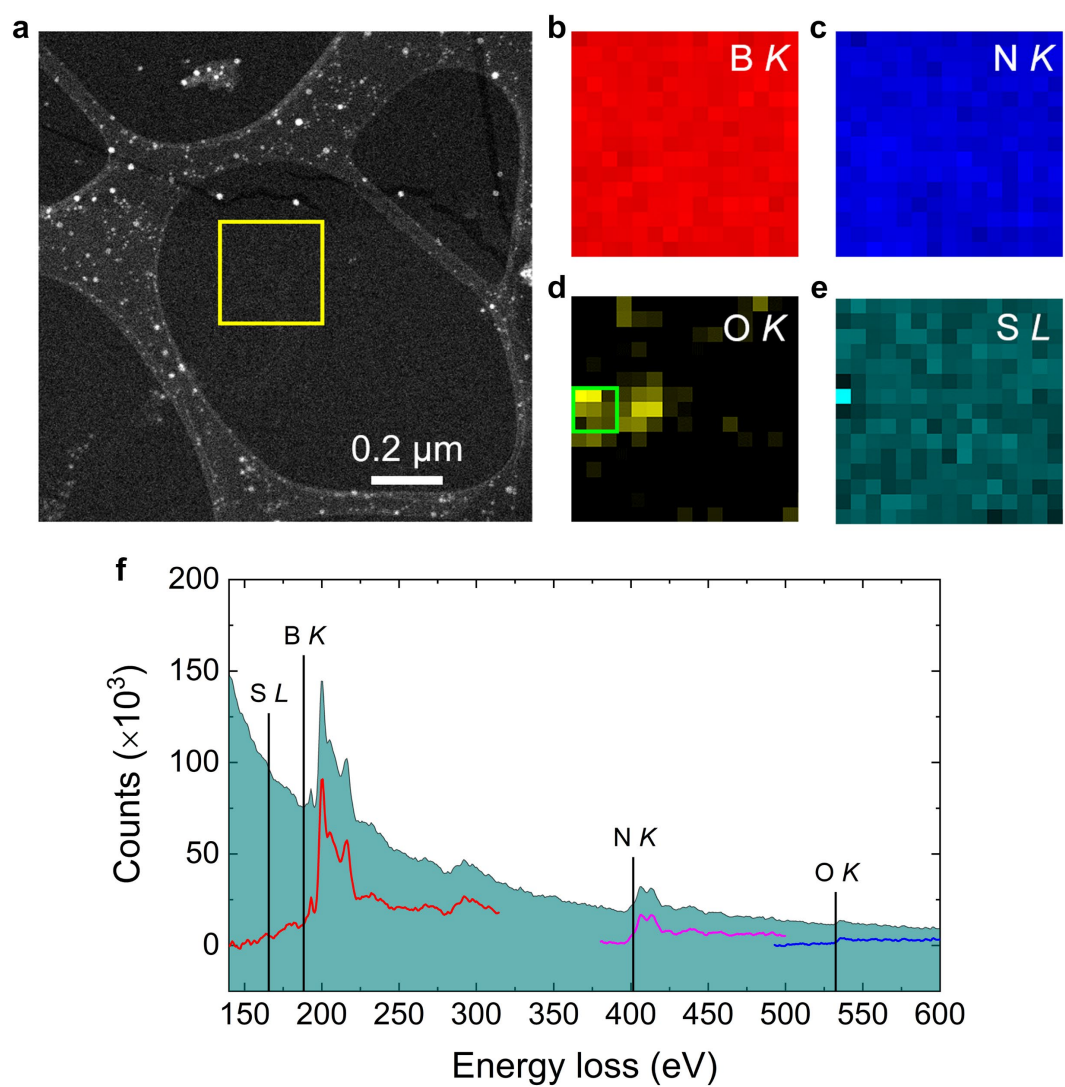

**Figure S10.** a) ADF-STEM image for the *h*-BN encapsulated WS₂ crystals. b-e) The EELS maps for the boron (b), nitrogen (c), oxygen (d) *K*-edge, and the sulfur (e) *L*-edge measured at the region marked by the yellow square box in (a). f) EELS spectrum measured at the region indicated by the green square box in (e).

## S11. Theoretical calculation for EELS analysis

**DFT calculation** We employee the first-principles density functional theory implemented in the full-potential linearized plane wave (FLAPW) with local orbitals with the ELK code.[9] In the calculation, the local-density approximation exchange correlation functional is adopted and size of basis set, rgkmax, is set to 7.0. $6 \times 6 \times 1$ **k**-point grids are considered for WS$_2$ (2, 2) supercell with lattice constant of 3.13 Å.[10,11] To consider core-hole excitation state, we perform the pseudo core-hole approach for a local orbital of orbital angular momentum $l = 0$, i.e., for the $K$-shell, in the FPLAPW with deep linearization energy (–25 atomic unit). The pseudo core-hole is made of one of the oxygen nuclei becomes positively charged (+1e) in the all-electron potential together with an additional electron (–1e) simultaneously created at the valence Kohn-Sham orbitals. After the first-principles self-consistent calculation, interactions between all the Kohn-Sham states in the unit-cell and the core-hole become properly demonstrated. The notation of Kohn-Sham Hamiltonian $\widehat{H}_{ks}[\rho]$ and eigenstates $|\mathbf{k}, n\rangle$ with core-hole of oxygen $K$-edge can be rewritten as $\widehat{H}_{ks}[\rho; O_i]|\mathbf{k}, n; O_i\rangle = E_{n,\mathbf{k}}|\mathbf{k}, n; O_i\rangle$. The $|\mathbf{k}, n; O_i\rangle$ stands for the Kohn-Sham eigenket at a band index $n$ with $\mathbf{k}$ momentum. The $O_i$ indicates that the DFT calculation is performed under the Hamiltonian including a core-hole is created at given $i$th oxygen atom of the oxygen molecule ($O_2$: $O_{i=1}O_{i=2}$).

**EELS calculation and Orbital distribution** The EELS can be directly obtained via the following relation,

$$\epsilon_2(\omega; O_i) \sim \sum_{n,m,\mathbf{k}} f_{\mathbf{k},m}(2 - f_{\mathbf{k}+\mathbf{q},n})|\langle \mathbf{k} + \mathbf{q}, n; O_i|e^{i\mathbf{q}\cdot\mathbf{r}}|\mathbf{k}, m; O_i\rangle|^2 \delta(E_{n,\mathbf{k}+\mathbf{q}} - E_{m,\mathbf{k}} - \omega)$$

The $f_{\mathbf{k},n}$ represents occupation number of electrons. In the above relation, the $|\mathbf{k}, m; O_i\rangle$ should be an oxygen $K$-shell state of the $i$th oxygen atom. We define the direction of unit $\mathbf{q}$ is approximately aligned along the in-plane cell vectors. The orbital distribution corresponding to each EELS peak can be extracted by following relation, $\rho(\mathbf{r}, O_i)_E = \sum_n |\langle \mathbf{r}|\mathbf{k}, n; O_i\rangle|^2 \delta(E - E_{n,\mathbf{k}} - E_K)$, here $E_K$ is the energy of given core-hole $K$-shell of $O_i$ with constant scissor operator. Note, in the Figure 2(b) of the manuscript, the x-ray absorption for physisorbed oxygen molecules provides identical transition pathways for $\epsilon_2(\omega; O_{i=1})$ and $\epsilon_2(\omega; O_{i=2})$ equivalently so that $\rho(\mathbf{r})_E \to \rho(\mathbf{r}; O_{i=1})_E$ and $\epsilon_2(\omega) \to \epsilon_2(\omega; O_{i=1})$ are displayed. For the chemisorded oxygen molecules in the Figure 2(c) of the manuscript, however, each oxygen is not equivalent to each other. Thus we show the averaged EELS including two number of cases for core-hole excitations, say, $\epsilon_2(\omega) \to \epsilon_2(\omega; O_{i=1}) + \epsilon_2(\omega; O_{i=2})$ and $\rho(\mathbf{r})_E \to \rho(\mathbf{r}; O_{i=1})_E$. We will provide the $\rho(\mathbf{r}; O_{i=2})_E$ in the Figure S11.

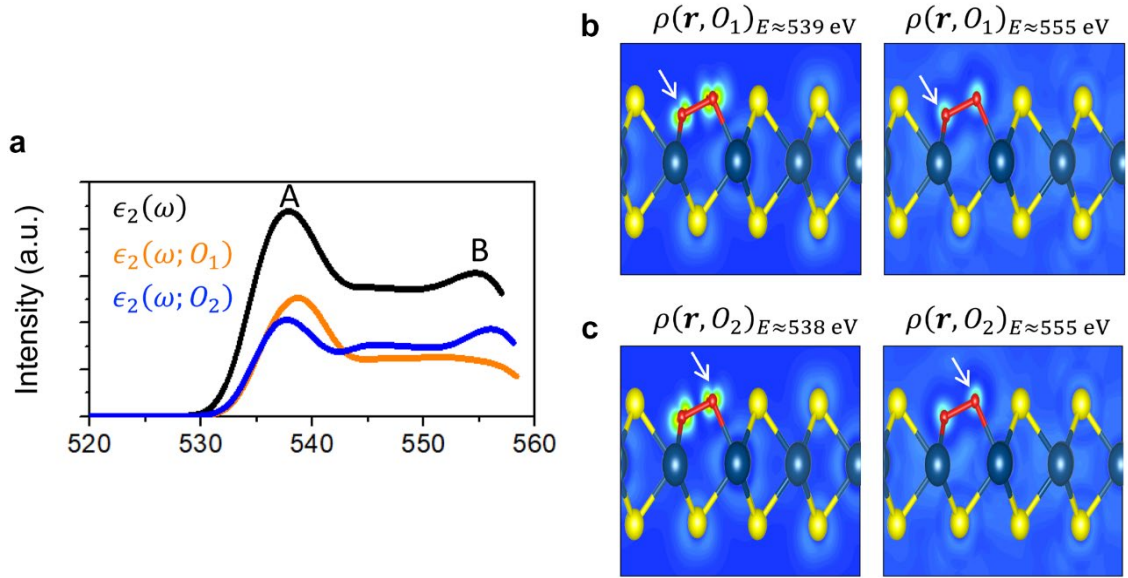

**Figure S11.** Core-hole dependent EELS spectra and orbital distributions. As long as chemisorbed, the two oxygens are not equivalent to each other and oxygen-dependent core-hole states become distinguishable. Here we assume that the experimental EELS is equally induced from the two kinds of oxygen core-holes. a) core-hole decomposed EELS, $\epsilon_2(\omega) = \epsilon_2(\omega; O_1) + \epsilon_2(\omega; O_2)$, and the $\epsilon_2(\omega)$ is displayed in the Figure 2(c) of the main text. b,c) The $\rho(r, O_i)$ for each peak, A and B, are displayed. White arrows indicate the oxygen atom with core-hole, i.e., $O_{i=1}$ (b) and $O_{i=2}$ (c). The peak A (black) at 538 eV can be decomposed into the two independent peaks at 537.7 eV (blue) and 538.7 eV (yellow). The physical origin of these decomposed peaks is found to be similar as depicted in the (b) and (c). However, the peak B at 555 eV is mostly contributed from that of the $\epsilon_2(\omega; O_2)$.

## S12. Power-law for the neutral exciton emission intensity at room temperature

The exciton generation and recombination in the steady-state photoluminescence can be described by the following rate equation:[12]

$$G = \frac{n_X}{\tau_X} + T n_X n_e + A n_X n_e + \gamma n_X^2 \qquad (1)$$

where $G$ is the exciton generation rate, $n_X$ is the neutral exciton density, $\tau_X$ is the exciton lifetime, $n_e$ is the electron density, $T$ is the trion formation coefficient, $A$ is the exciton-electron Auger coefficient, and $\gamma$ is the exciton–exciton annihilation coefficient. Note that the contribution of localized excitons and dark excitons can be neglected by the thermal activation effect at room temperature.[13,14] The exciton-electron Auger ($A n_X n_e$) and exciton-exciton annihilation ($\gamma n_X^2$) processes can be neglected at a low level of excitation, and thus the exciton-to-trion conversion process becomes a dominant nonradiative decay at our experimental conditions. When the neutral exciton and free electron density increase with increasing the excitation power density (for the bare WS$_2$), the equation (1) can be simplified as the following equation:

$$G \cong T n_X n_e \qquad (2)$$

When the free electron density increases with the power dependence of $P^\mu$ ($n_e \sim P^\mu$), the neutral exciton density can be written by

$$n_X \cong \frac{G^{1-\mu}}{T} \qquad (3)$$

where the exciton generation rate is proportional to the excitation power ($G \propto P$). The neutral exciton emission intensity for the excitation power can be expressed as the following relation (4), showing that the neutral exciton emission intensity follows the power-law of $P^{1-\mu}$.

$$PL = \frac{n_X}{\tau_X} \cong \frac{G^{1-\mu}}{\tau_X T} \propto P^{1-\mu} \qquad (4)$$

**S13. Neutral exciton-to-trion conversion in the bare and *h*-BN encapsulated WS$_2$**

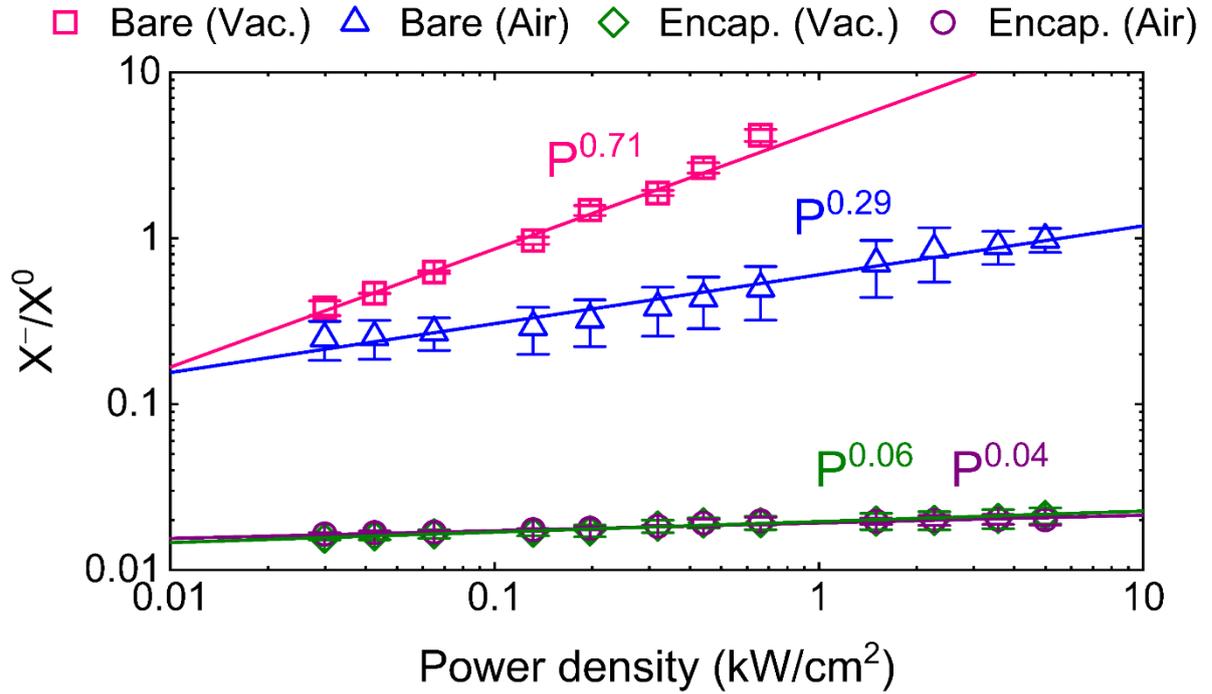

**Figure S13.** Photoluminescence intensity ratio (X$^-$/X$^0$) of the trion to the neutral exciton with increasing the excitation power density. In a power-law ($X^-/X^0 \propto P^\mu$), for the bare WS$_2$, the neutral excitons are converted to the trions in proportional to $P^\mu$ with the exponent $\mu$ of 0.71 and 0.29 in the ambient vacuum and air conditions with increasing the excitation power, while showing the exponent of 0.06 and 0.04 for the *h*-BN encapsulated WS$_2$ in the vacuum and air ambient conditions, respectively. As shown in Figures 3(a,b), the exciton emission intensity in the bare and *h*-BN encapsulated WS$_2$ increases with the exponent $\alpha \cong 1 - \mu$ under the vacuum and air conditions, indicating that the neutral exciton-to-trion conversion is the predominant process in the nonradiative decay of the excitons. Thus, Auger recombination process can be neglected under our excitation conditions.

## S14. Estimation of the free electron density using mass action law

To determine the free electron density in the bare and $h$-BN encapsulated WS$_2$, we used the mass action law describing the formation of the trion ($X^-$) from the neutral exciton ($X^0$) and the free electrons ($e$). The formation of trion can be expressed by the following chemical equation:[15,16]

$$X^0 + e \Leftrightarrow X^- \tag{1}$$

From the chemical equation, the density of each carrier can be related by

$$\frac{n_{X^-}}{n_{X^0}} = \frac{n_e}{K(T)} \tag{2}$$

where $n_{X^-}$, $n_{X^0}$, and $n_e$ correspond to the density of trions, neutral excitons, and electrons. $K(T)$ is the temperature-dependent equilibrium constant defined by the following equation:

$$K(T) = \left(\frac{4 m_{X^0} m_e}{\pi \hbar^2 m_{X^-}}\right) k_B T \exp\left(-\frac{E_b^{X^-}}{k_B T}\right) \tag{3}$$

where $m_{X^-}$, $m_{X^0}$, and $m_e$ are the effective mass of the trion, neutral exciton, and electron. The effective mass of the trion and neutral exciton was calculated to be $1.33 m_0$ ($m_{X^-} = 2 m_e + m_h$) and $0.89 m_0$ ($m_{X^0} = m_e + m_h$), where $m_0$ is the electron mass, $m_e$ ($0.44 m_0$) is the effective mass of the electron, $m_h$ ($0.45 m_0$) is the effective mass of the hole, $k_B$ is the Boltzmann constant, $T$ is the temperature, and $E_b^{X^-}$ is the trion binding energy ($\sim$34 meV).[17,18]

By using the equations (2) and (3), we can obtain the following parameter:

$$\frac{n_{X^0} n_e}{n_{X^-}} = \left(\frac{4 m_{X^0} m_e}{\pi \hbar^2 m_{X^-}}\right) k_B T \exp\left(-\frac{E_b^{X^-}}{k_B T}\right) = 3.41 \times 10^{12} \ cm^{-2} \tag{4}$$

The intensity weight ($I_{X^-}/I_{Total}$) of trions in the photoluminescence spectra measured from the bare and $h$-BN encapsulated WS$_2$ crystals can be written by the following equation:

$$\frac{I_{X^-}}{I_{total}} = \frac{\gamma_{X^-} n_{X^-}}{\gamma_{X^0} n_{X^0} + \gamma_{X^-} n_{X^-}} = \frac{\frac{\gamma_{X^-}}{\gamma_{X^0}} \frac{n_{X^-}}{n_{X^0}}}{1 + \frac{\gamma_{X^-}}{\gamma_{X^0}} \frac{n_{X^-}}{n_{X^0}}} \tag{5}$$

where $\gamma_{X^0}$ and $\gamma_{X^-}$ are the decay rate of neutral excitons and trions, respectively. $\gamma_{X^-}/\gamma_{X^0}$ was estimated by measuring the lifetimes of neutral excitons and trions.

By comparing the parameter (4) and the equation (5), we can obtain the following equation:

$$n_e = 3.41 \times 10^{12} \frac{\left(\frac{\gamma_{X^0}}{\gamma_{X^-}}\right)\left(\frac{I_{X^-}}{I_{total}}\right)}{\left(1 - \frac{I_{X^-}}{I_{total}}\right)} \ [cm^{-2}] \tag{6}$$

From the photoluminescence spectra measured as a function of the excitation power density, we estimated the free electron density in the bare and *h*-BN encapsulated WS$_2$ crystals.

## S15. Quantitative estimate for the number of desorbed/adsorbed oxygen molecules on sulfur vacancies in-between the vacuum and air envrionments

To investigate a quantitative estimate of the adsorption/desorption of oxygen molecules on WS$_2$, we fabricated the bare WS$_2$ capacitor device (Figure S15a), enabling the electrostatic control of electron cencentration as a function of gate voltage ($V_g$). The gate-voltage-dependent PL measurements were performed at an excitation power density of 0.196 kW/cm$^2$. As shown in Figures S15c and S15d, the charge neutral poins of the bare WS$_2$ device were determined at $V_g \cong -14\ V$ and $V_g \cong -9\ V$ under the vacuum (Figure S15c) and air (Figure S15d) ambient conditions, respectively. We calculated the electron densities in the bare WS$_2$ device at the vacuum and air environments using the equation $\Delta n_{WS_2} = C \times (V_g - V_{neutral})/e$, where $\Delta n_{WS_2}$ indicates the electron density injected by the gate voltage, $C$ is the capacitance of the used $h$-BN layer (6.55 × 10$^{-4}$ F·cm$^{-2}$), $V_g$ is the gate voltage, $V_{neutral}$ is the onset voltage determined by the neutral point, and $e$ is the electronic charge.[19] The electron densities were estimated to be 1.15 × 10$^{13}$ cm$^{-2}$ (vacuum) and 3.69 × 10$^{12}$ cm$^{-2}$ (air) at $V_g = 0\ V$, respectively, while those estimated using mass-action law were 1.11 × 10$^{13}$ cm$^{-2}$ and 2.29 × 10$^{12}$ cm$^{-2}$ at the vacuum and air environments, respectively, showing a good agreement. In-between the vacuum and the air environments, the free electron density induced by the desorption of oxygen molecules is estimated to be 7.77 × 10$^{12}$ cm$^{-2}$ by considering the difference in the free electron densities at the vacuum and air environments. Since an oxygen molecule gains 1.083 electrons from WS$_2$,[2] the number of desorbed/adsorbed oxygen molecules on the sulfur vacanies were estimated to be 7.17 × 10$^{12}$ cm$^{-2}$.

On the other hand, we also fabricated the $h$-BN encapsulated WS$_2$ capacitor devices (Figure S15b). As shown in Figures S15e and S15f, the charge neutral points of the $h$-BN encapsulated WS$_2$ device were determined at $V_g \cong 0.5\ V$ for both the vacuum and air ambient conditions, resulting in the electron densities of 2.05 × 10$^{11}$ cm$^{-2}$.

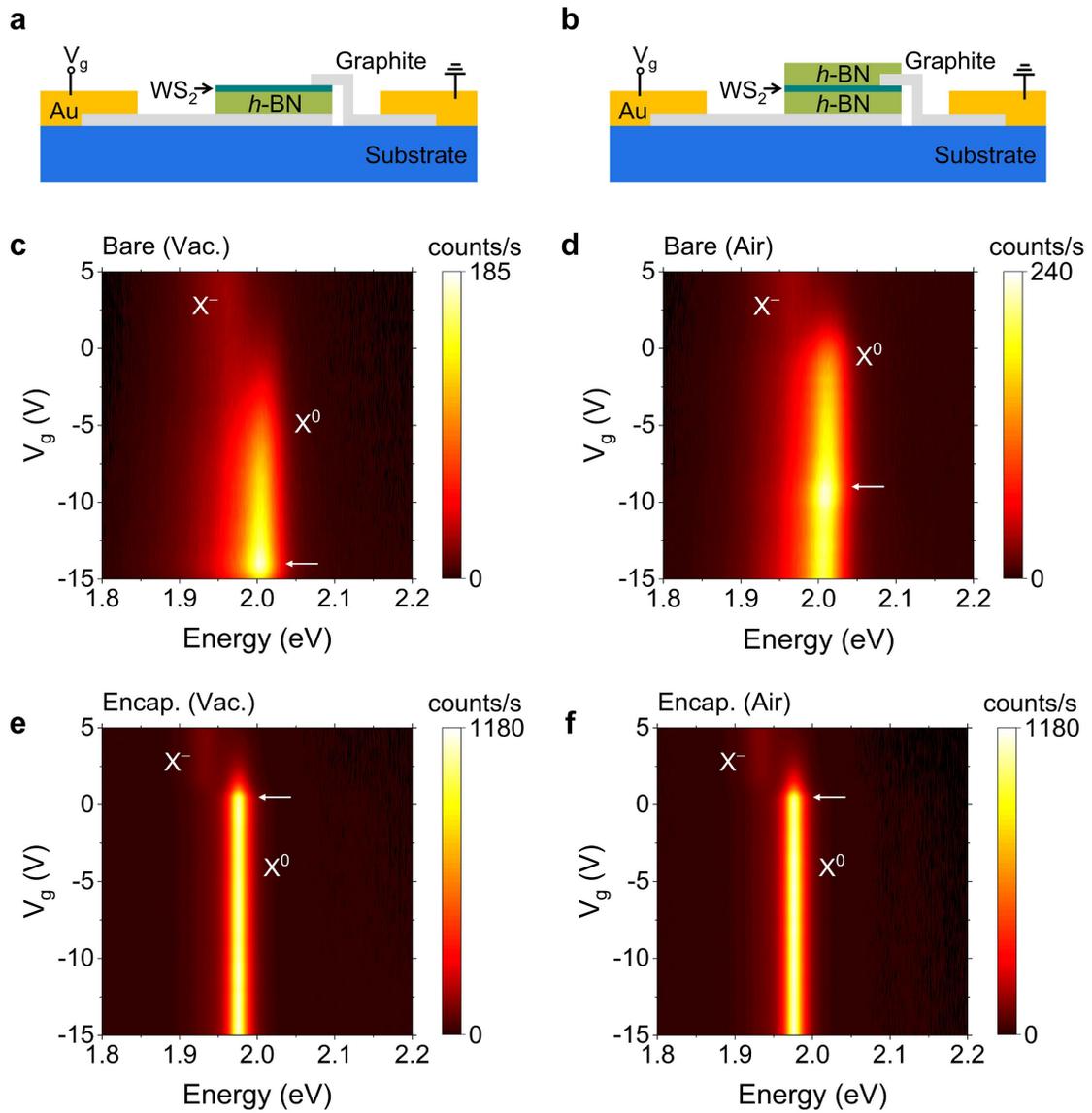

**Figure S15.** a,b) Schematic illustration for the bare (a) and *h*-BN encapsulated (b) WS$_2$ capacitor devices. c,d) Gate-voltage-dependent photoluminescence spectral maps for the bare WS$_2$ device under the vacuum (c) and air (d) ambient conditions. e,f) Gate-voltage-dependent photoluminescence spectral maps for the *h*-BN encapsulated WS$_2$ device under the vacuum (e) and air (f) ambient conditions.

**S16. Photoluminescence decay curves of the neutral excitons under the vacuum ambient condition**

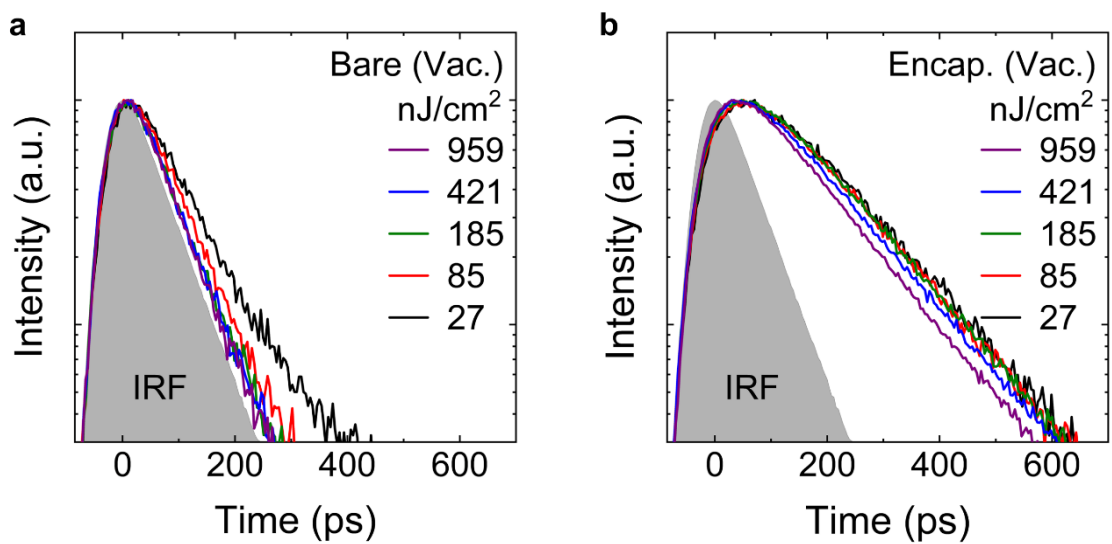

**Figure S16.** a,b) Photoluminescence decay curves of the neutral excitons measured as a function of the excitation power density for the bare (a) and *h*-BN encapsulated (b) WS$_2$ crystals under the vacuum ambient condition.

## S17. Lifetime of excitons for the bare and *h*-BN encapsulated WS$_2$ measured as a function of the energy fluence

To extract the lifetime of excitons measured in the vacuum and air ambient conditions for the bare and *h*-BN encapsulated WS$_2$, the time-resolved photoluminescence decay curves are fitted using an exponential model based on instrument response function (IRF) defined as the following equation:

$$I(t) = \int_{-\infty}^{t} IRF(t') A_1 e^{-\frac{t-t'}{\tau_1}} dt$$

where $A_1$ and $\tau_1$ are the amplitude and the lifetime for the single exciton component.

**Table S1.** The lifetime of the neutral excitons measured as a function of the energy fluence for the bare WS$_2$ under the vacuum and air ambient conditions.

| Energy fluence (nJ/cm$^2$) | Vacuum (bare) | Air (bare) |
|---|---|---|
| 27 | 49 ps | 74 ps |
| 85 | 37 ps | 63 ps |
| 185 | 33 ps | 57 ps |
| 421 | 27 ps | 47 ps |
| 959 | 19 ps | 43 ps |

**Table S2.** The lifetime of the neutral excitons measured as a function of the energy fluence for the *h*-BN encapsulated WS$_2$ under the vacuum and air ambient conditions.

| Energy fluence (nJ/cm$^2$) | Vacuum (encap.) | Air (encap.) |
|---|---|---|
| 27 | 139 ps | 136 ps |
| 85 | 136 ps | 133 ps |
| 185 | 132 ps | 129 ps |
| 421 | 126 ps | 123 ps |
| 959 | 116 ps | 113 ps |

**S18. Exciton lifetime as a function of free electron density and annihiliation rate constant in the *h*-BN encapsulated WS$_2$ capacitor devices**

**Table S3.** The lifetime of the neutral excitons in the *h*-BN encapsulated WS$_2$ capacitor device measured as a function of the gate voltage at a fixed pump fluence of 27 nJ/cm$^2$.

| Gate voltage (V) | Injeced electron density (cm$^{-2}$) | Lifetime (ps) |
|---|---|---|
| 0.5 | 2.05 × 10$^{11}$ | 139 |
| 0.6 | 2.46 × 10$^{11}$ | 99 |
| 0.7 | 2.87 × 10$^{11}$ | 62 |
| 0.8 | 3.28 × 10$^{11}$ | 45 |
| 0.9 | 3.69 × 10$^{11}$ | 36 |

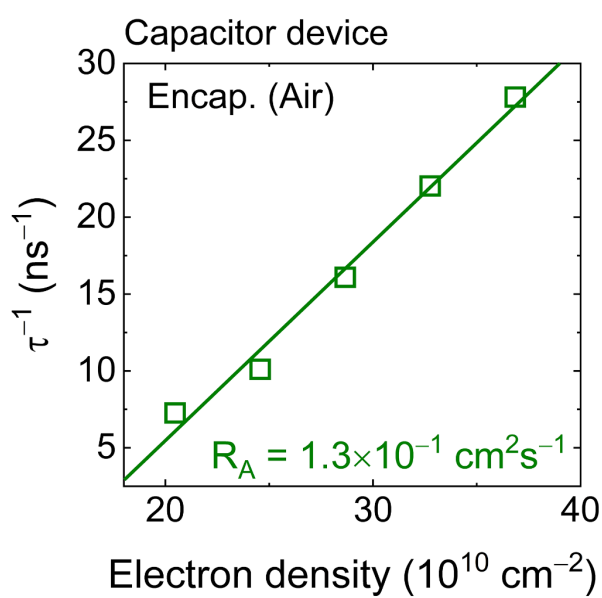

**Figure S18.** Recombination rate ($\tau^{-1}$) as a function of free electron density for the *h*-BN encapsulated WS$_2$ capacitor devices. A linear fit for the $\tau^{-1}$ results in the exciton annihilation rate constant ($R_A$) due to the exciton-to-trion conversion process.

**S19. Gate-voltage-dependent phtoluminescence decay curves of the neutral excitons in the *h*-BN encapsulated WS$_2$ capacitor devices**

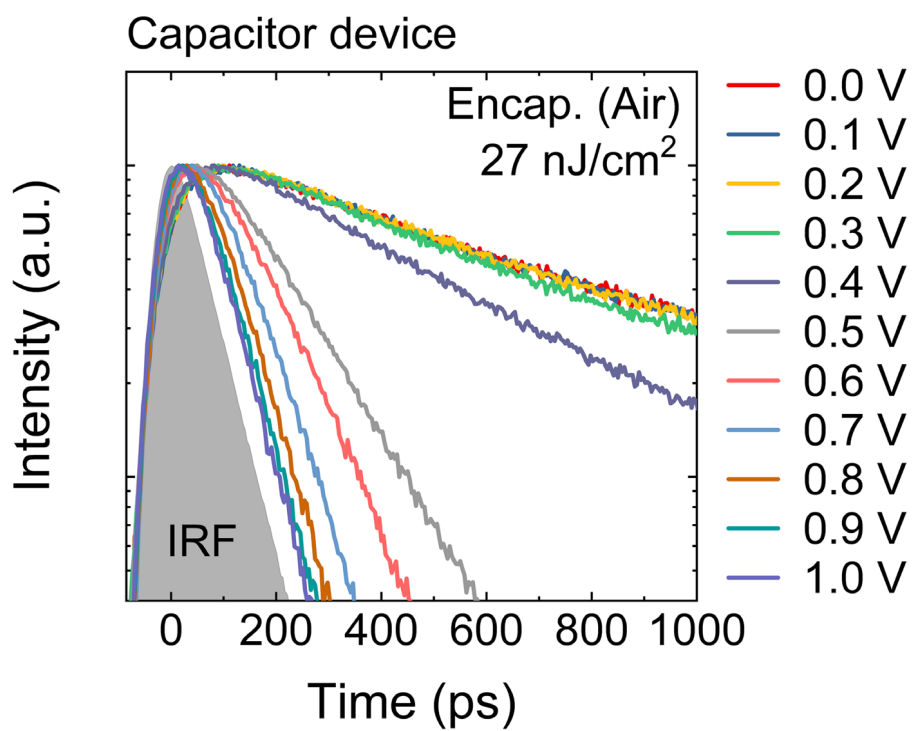

**Figure S19.** Gate-voltage-dependent photoluminescence decay curves of the neutral excitons in the *h*-BN encapsulated WS$_2$ capacitor devices under the air ambient condition.